\documentstyle [eqsecnum,preprint,aps,epsf,floats,fixes]{revtex}
\tightenlines
\begin {document}

\def\hard{{\rm hard}}
\def\D{{\bf D}}

\def\E{{\bf E}}

\def\B{{\bf B}}

\def\A{{\bf A}}

\def\J{{\bf J}}
\def\x{{\bf x}}

\def\p{{\bf p}}
\def\k{{\bf k}}
\def\q{{\bf q}}
\def\v{{\bf v}}
\def\u{{\bf u}}
\def\g{{\rm g}}
\def\grad{\mbox{\boldmath$\nabla$}}
\def\noise{\zeta}
\def\bnoise{\mbox{\boldmath$\noise$}}

\def\CA{C_{\rm A}}
\def\nf{n_{\rm f}}
\def\pl{{\rm pl}}
\def\taularge{\tau_{\rm large}}
\def\tauany{\tau_{\rm small}}
\def\tM{t_{\rm M}}
\def\I{{\cal I}}
\def\R{{\cal R}}
\def\cD{{\cal D}}
\def\eff{{\rm eff}}
\def\tr{{\rm tr}}
\def\TR{t_\R}
\def\TRp{t_{\R'}}
\def\eq{{\rm eq}}

\def\PiT{\Pi_{\rm T}}
\def\typespin{{\hbox{\tiny type}\atop\hbox{\tiny spin}}}

\def\V{{\cal V}}
\def\VR{{\cal V}_{\rm R}}
\def\eps{\epsilon}

\def\Dt{\Delta t}

\def\half{{\textstyle{1\over2}}}
\def\fourth{{\textstyle{1\over4}}}

\preprint {\vbox{\hbox {~UW/PT 98--10}\hbox{MIT CTP--2779}}}

\title  {Hot B violation, color conductivity, and log(1/\boldmath$\alpha$)
effects}

\author {Peter Arnold}

\address
    {%
    Department of Physics,
    University of Virginia,
    Charlottesville, VA 22901
    }%
\author{Dam T. Son}
\address
    {%
    Center for Theoretical Physics,
    Department of Physics, \\
    Massachusetts Institute of Technology,
    Cambridge, MA 02139
    }%
\author{Laurence G. Yaffe}
\address
    {%
    Department of Physics,
    University of Washington,
    Seattle, Washington 98195
    }%
\date {August 1998}

\maketitle
\vskip -20pt

\begin {abstract}%
{%
B\"odeker has recently argued that non-perturbative processes in very high
temperature non-Abelian plasmas (such as electroweak baryon number violation
in the very hot early Universe) are logarithmically enhanced over previous
estimates and take place at a rate per unit volume of order $\alpha^5 T^4
\ln(1/\alpha)$ for small coupling.  We give a simple physical interpretation
of B\"odeker's qualitative and quantitative results in terms of Lenz's
Law---the fact that conducting media resist changes in the magnetic
field---and earlier authors' calculations of the color conductivity of such
plasmas.  In the process, we resolve some confusions in the literature about
the value of the color conductivity and present an independent calculation.
We also discuss the issue of whether the classical effective theory proposed
by B\"odeker has a good continuum limit.
}%
\end {abstract}

\thispagestyle{empty}


\section {Introduction}

Standard electroweak theory
violates baryon number via non-perturbative processes involving
the electroweak anomaly.\footnote{
   For some reviews of electroweak baryon number violation and electroweak
   baryogenesis, see ref.\ \cite{reviews}.
}
Such processes are exponentially suppressed
under normal conditions, but are unsuppressed at very high
temperatures in the early Universe.
Non-perturbative baryon number violation is a key
ingredient in scenarios for electroweak baryogenesis, which attempt
to explain the matter/anti-matter asymmetry of the Universe in terms
of the physics of the electroweak phase transition.  Such scenarios
typically depend on (among other things) the equilibrium rate of
baryon number violation in the hot, symmetry-restored phase of
electroweak theory.%
\footnote{
   We use the term ``symmetric phase'' loosely since,
   depending on the details of the Higgs sector, there may not be any
   sharp transition between the symmetric and ``symmetry-broken'' phases
   of the theory \cite{notransition}.  A sharp transition is in fact
   required for electroweak baryogenesis.  The analysis of this paper
   applies directly whenever the temperature is sufficiently high that
   the infrared dynamics of the Higgs is irrelevant at lengths
   of $O(1/g^2 T)$, which is the case either (a) far above the electroweak
   phase transition or ``crossover,'' or (b) in the symmetric phase at
   the transition in cases where there is a first-order transition and
   the transition is not exceedingly weak.
}
The rate of baryon number violation---and more generally the rate of
any generic non-perturbative process in high-temperature non-Abelian
plasmas---has long been a source of theoretical confusion.  In fact,
it is only recently becoming clear how the rate scales with the
fine structure constant $\alpha$ of the relevant gauge interactions
in the arbitrarily weak coupling limit.

For non-Abelian plasmas at ultra-relativistic
temperatures, 
non-perturbative fluctuations of the gauge field
are associated with magnetic fluctuations over distance scales
$R \sim 1/g^2 T$ (to be reviewed momentarily).
For a long time in the literature, it was assumed that the time scale $t$ of
non-perturbative processes was also of order $1/g^2 T$, and that the
rate $\Gamma$ per unit volume was therefore of order
$1/L^3 t \sim \alpha^4 T^4$.  Two years ago, we argued \cite{ASY1} that damping
effects
in the plasma slow the time scale down to $t \sim 1/g^4 T$, giving a rate of
$\Gamma \sim 1/L^3 t \sim \alpha^5 T^4$.
(See also refs.~\cite{Huet&Son,Arnold}.)
More recently, B\"odeker \cite{Bodeker} has claimed that there is an additional
logarithmic suppression of the time scale, so that
\begin {equation}
   t \sim {1 \over g^4 T \ln(1/g)} \,,
   \qquad\qquad \hbox{and} \qquad\qquad
   \Gamma \sim \alpha^5 T^4 \ln(1/\alpha) \,.
\end {equation}
B\"odeker has also proposed an effective theory for the relevant distance
and time scales in the form of simple stochastic dynamics for the gauge
fields.  Numerical simulation of this effective theory would give the
non-perturbative numerical coefficient $c$ of the logarithm:
\begin {equation}
   \Gamma \simeq c \, \alpha^5 T^4 \ln(b/\alpha)
\end {equation}
for small $\alpha$.  (No one has yet proposed an explicit way to calculate
the constant $b$ under the log, and one should expect there to be sub-leading
corrections suppressed only by powers of $1/\ln \alpha$.)
The goal of the present work is to show that B\"odeker's results can be
reproduced and interpreted through a simple argument based on the fact
that plasmas are conductive.

Before presenting the essential argument, let us take a moment to review the
physical origin of the length scale $1/g^2 T$ associated with non-perturbative
fluctuations.  (For more formal arguments, see \cite{scale}.)
Imagine a fluctuation of the gauge field of spatial size $R$
and amplitude $A$.  Non-perturbative means that, for example, $gA$ is not a
perturbation in the covariant derivative $\D = \grad - i g\A$.  So
non-perturbative means $A \gtrsim O(1/gR)$, and hence the energy $E$ of this
fluctuation is $\gtrsim O(1/g^2 R)$.  The probability of such a fluctuation in
energy is exponentially suppressed by the Maxwell-Boltzmann factor
$\exp(-\beta E) \sim \exp[-1/(g^2 RT)]$
unless $R \gtrsim O(1/g^2 T)$.  Because of entropy effects,
non-perturbative processes will be dominated by the smallest size scale
for which the probability is unsuppressed
(since there are more small-wavelength
degrees of freedom than large-wavelength ones), and so the characteristic
length scale of non-perturbative physics is $R \sim 1/g^2 T$.
Static electric fields are screened by the Debye effect on smaller distance
scales, of order $1/g T$.  For this reason, non-perturbative physics
in the hot plasma is essentially magnetic.  More technically, it is only
the transverse degrees of freedom of the gauge field which are important.

In the next section, we present the simple relationship between the color
conductivity and non-perturbative dynamics at leading-log order, and reproduce
B\"odeker's effective theory for the non-perturbative dynamics.  In section 3,
we review the somewhat confusing literature on color conductivity and present
our own calculation based on the Boltzmann equation with a collision term.
Finally, in section 4, we argue that B\"odeker's
effective theory is ultraviolet insensitive---a crucial property for numerical
simulations.


\section {The Essential Argument}
\label{sec:essential}

We now turn to the essence of the argument, which is quite short.
It is based on realizing that the dynamics of magnetic fluctuations
in plasmas is slowed down by Lenz's law: conducting media resist changes
in magnetic field.  In the context of high temperature baryon number
violation, this qualitative explanation of the slow time scale for
non-perturbative processes is due to Guy Moore \cite{Moore}.  Let's make it
quantitative.  This derivation will be a little fast and loose, but its
advantage is that the physics is very simple.

Imagine splitting
the gauge field into soft degrees of freedom---those associated with
momenta of order $g^2 T$, and hard degrees of freedom---those associated
with much higher momenta such as $T$.
The details of exactly how this split is made
will not be relevant at the order we shall consider.%
\footnote{%
   That's fortunate, because trying to make such a split explicit creates
   a host of difficulties.  See ref.~\cite{mess}.%
}
The amplitude of fluctuations is non-perturbative for the soft modes
but perturbative for the hard ones.
As is well known \cite{classical},
the soft modes are also effectively classical---there are
a large number of quanta in each mode because of Bose statistics.
Now, treating the soft modes classically, start with the Maxwell equation
\begin {equation}
   \D \times \B = D_t \, \E + \J_\hard
\label{eq:eq1}
\end {equation}
for the soft degrees of freedom, where $\B = \D \times \A$ and where all
covariant derivatives are to be understood as only involving the soft
gauge field degrees of freedom.
$\J_\hard$ is the color current%
\footnote
    {%
    We are using ``color'' as a descriptive name for
    some non-Abelian gauge field.
    It should be emphasized that all discussion of ``color''
    is applicable to the dynamics of,
    in particular, the SU(2) electroweak gauge field.
    }
due to the hard degrees of freedom, which we shall later see is
dominated by excitations with momenta of order $T$.
It is important to distinguish between the hard momenta of the particles
which contribute to $\J_\hard$ and the momentum components of $\J_\hard$
itself (which is bilinear in the fundamental fields).  It is the
soft momentum components of $\J_\hard$ which are relevant in the context
of (\ref{eq:eq1}).

Plasmas are conductors.
Hence, for sufficiently small momentum and frequency
(exactly how small will be discussed later), we have
\begin {equation}
   \J_\hard = \sigma \E \,,
\end {equation}
where $\sigma$ is the color analog of conductivity.  The Maxwell equation
then becomes
\begin {equation}
   \D \times \B = D_t \, \E + \sigma \E
   .
\label{eq:eq2}
\end {equation}
Let us assume that non-perturbative processes will be slow enough (which
we will verify {\it a posteriori}) that we can neglect the time derivative
term.  Then the Maxwell equation becomes simply
\begin {equation}
   \D \times \B = \sigma \E
   .
\end {equation}
In $A_0 = 0$ gauge, this is a simple first-order equation of motion:
\begin {equation}
   \sigma {d\over dt} \, \A = -\D \times \B
   .
\label{eq:eq3}
\end {equation}

This equation is dissipative and describes the relaxation of fluctuations of
the soft fields away from equilibrium.
The dissipation results from interactions
of the soft modes with the hard degrees of freedom,
which are accelerated by and steal energy from the soft fields.
Interactions with the hard modes, however, not only
provide dissipation for the soft modes; they also serve as a source of
thermal noise.
In the above analysis,
the noise has been implicitly disregarded, and we will need to put it
in if we wish to describe equilibrium fluctuations.
Fortunately, this is simple to do after the fact because noise and dissipation
are intimately related by the fluctuation-dissipation theorem.
In the language of an effective theory of the soft modes, equilibrium
requires a delicate
balance between the soft modes' excitation from thermal noise and their
dissipative decay.

To be more specific, note that (\ref{eq:eq3}) has the general form
\begin {equation}
   \sigma {d\over dt} \, \q = - \grad_\q V(\q) ,
\end {equation}
where
$V(\q)$ is the potential energy of the degrees of freedom $\q$
(which in our case is the non-Abelian magnetic energy
${1\over2} \int_\x B^2$).
Such systems are common in physics, and a simple way to incorporate
thermal noise is to include a random force $\bnoise$:
\begin {equation}
   \sigma {d\over dt} \, \q = - \grad_\q V(\q) + \bnoise .
\label{eq:toy}
\end {equation}
This is a typical example of a Langevin equation.
The simplest possible choice of thermal noise,
Gaussian white noise,
reproduces the correct equilibrium distribution $\exp(-\beta V)$ if
the noise variance is suitably scaled with the amount of dissipation,
\begin {equation}
   \langle \noise_i(t) \noise_j(t') \rangle =
   2 \sigma \, T \, \delta_{ij} \, \delta(t{-}t') \,.
\label{eq:toynoise}
\end {equation}
This well-known result can be verified by converting the
Langevin equation (\ref{eq:toy}) into a Fokker-Planck equation for the
probability distribution.
(See, for example, chapter 4 of ref.\ \cite{ZinnJustin}.)

Why should one believe the noise distribution is so simple?
First, the noise can be treated as Gaussian
if the soft dynamics of interest has a time scale large
compared to the decorrelation time of the noise,
which is caused by fluctuations of the hard modes.
Averaging the noise over time scales small compared to the soft dynamics
scale but large compared to the noise decorrelation time, the central limit
theorem implies that the resulting distribution will approach a Gaussian
shape.
We will see later (sec.\ \ref{sec:roughsig}) that in our case the relevant
decorrelation time for hard fluctuations is $1/(g^2 T \ln g^{-1})$ whereas
the time scale for soft dynamics is the longer scale $1/(g^4 T \ln g^{-1})$
asserted earlier.
Second, if the theory were linearized, then the fact that the spectrum
of this Gaussian noise is white noise would follow rigorously from the
fluctuation-dissipation theorem.  More generally, any noise spectrum
$f(\omega)$ may be regarded as frequency-independent
({\em i.e.}, white noise) at sufficiently small frequency $\omega$
provided $f(0)$ is finite and non-zero.
So effective theories for long time scales can generally be expected to have
Gaussian white noise.  Finally, one might wonder why there could not be some
non-linear coupling to the noise, in the form of a function $e(\q)$
multiplying the noise term $\bnoise$ in (\ref{eq:toy}).
Generically, the introduction of such a $\q$-dependence would
change the equilibrium distribution produced by (\ref{eq:toy}) so
that it would not correctly reproduce $\exp(-\beta V)$.

Based on the above discussion,
let us introduce noise as in (\ref{eq:toy}).
Translating back
to our particular system (\ref{eq:eq3}), we obtain the following effective
theory for the soft modes:
\begin{mathletters}%
\label{eq:eff}%
\begin {eqnarray}
   \sigma {d\over dt} \, \A &=& -\D \times \B + \bnoise ,
\label{eq:effa}
\\
   \langle \noise^a_i(t,\x) \, \noise^b_j(t',\x') \rangle^{\strut}
     &=& 2 \sigma \,T \,\delta^{ab} \delta_{ij}
     \,\delta(t{-}t') \,\delta(\x{-}\x'),
\end {eqnarray}%
\end{mathletters}%
where $i,j$ and $a,b$ are spatial vector and adjoint color indices,
respectively.
Those readers interested in a more technical
derivation of the noise term starting somewhat closer to first principles
should consult B\"odeker \cite{Bodeker}.

Astute readers may notice a peculiarity of (\ref{eq:eff}):
it introduces noise for
the longitudinal as well as transverse modes of $\A$,
whereas the effective theory is only meant to describe the transverse modes.
(The longitudinal modes are the pieces of $\E$ which contribute to
$\D\cdot\E$ and perturbatively correspond to polarizations
parallel to the spatial momentum $\k$.)
The noise-driven longitudinal dynamics generated by (\ref{eq:eff}) is nothing
more than a convenient fiction which simplifies the description of the
effective theory and which does
not affect the transverse dynamics of interest.
We shall henceforth ignore this issue in the present paper and
instead discuss it in detail elsewhere \cite{paper2}.


The effective equation (\ref{eq:eff}) turns out to have the wonderful
property that it is insensitive to how the soft modes are cut off
at large momentum.  (We will discuss this in greater depth in
section \ref {sec:uv}.)
It means that one can
ignore the soft/hard separation that was necessary to write (\ref{eq:eq1})
but which was never specified in detail.
It means that (\ref{eq:eff}) will be insensitive to short-distance
lattice cut-offs used in numerical simulations.
Finally, it also means that such simulations will not be plagued by
lattice artifacts, such as loss of rotational invariance, that were
thought to arise in other approaches \cite{Arnold}.

From (\ref{eq:effa}) and $\B = \D \times \A$, one can immediately see
that the time scale of non-perturbative dynamics is given by
\begin {equation}
   \sigma \, t^{-1} A \sim R^{-2} A ,
\label{eq:sigmaorder}
\end {equation}
so that
\begin {equation}
   t \sim R^2 \, \sigma \sim {\sigma \over g^4 T^2} .
\label{eq:tscale}
\end {equation}
Thus, one need only know the color conductivity $\sigma$.
There has been some confusion in the literature (described later)
about this quantity, but the correct value was first presented
by Selikhov and Gyulassy \cite{Selikhov}.
The color conductivity is of order
\begin {equation}
   \sigma \sim {T \over \ln(1/g)} \, .
\label{eq:roughsigma}
\end {equation}
We will review later how to understand this physically.
Inserting Eq.~(\ref {eq:roughsigma}) into
Eq.\ (\ref{eq:tscale}) then gives the time scale
\begin {equation}
    t \sim {1 \over g^4 T \ln(1/g)} \,,
\label{eq:tscale2}
\end {equation}
and so $\Gamma \sim \alpha^5 T^4 \ln(1/\alpha)$,
which has the logarithmic enhancement claimed by B\"odeker.
Later, we will see that earlier estimates \cite{ASY1} of the time scale
as $t \sim 1/(g^4 T)$ correspond to ignoring the effects of collisions on
the conductivity.
Note that
ignoring the time derivative term in (\ref{eq:eq2}) was justified since
the characteristic time scale (\ref {eq:tscale2}) is much greater than
the inverse conductivity $\sigma^{-1}$ determined by (\ref{eq:roughsigma}).

On a more quantitative level, the color conductivity is \cite{Selikhov}%
\footnote{
   The reader of ref.~\cite{Selikhov} should beware the final
   equation of that paper,
   eq.~(47).  In that equation, the authors replace their result
   by something rough and approximate.
}
\begin {equation}
    \sigma \approx {m_\pl^2 \over \gamma_\g} \,,
\label{eq:sigma}%
\end {equation}
where $m_\pl$ is the plasma frequency and
\begin {equation}
    \gamma_\g \approx \alpha \, \CA T \ln(1/g)
\label{eq:gamma}
\end {equation}%
is the damping rate for hard thermal gauge bosons \cite{gammag}.%
\footnote{%
\label{fn:width}%
   In the literature,
   the hard thermal ``damping rate'' is defined
   (in one-loop perturbation theory)
   as the imaginary part of the pole energy for a propagating gauge boson.
   In particular, it is defined so that the {\it amplitudes} of plasma
   waves decay as $\exp(-\gamma t)$.  This is in contrast to the standard
   usage
   of the ``width'' $\Gamma$ of a resonance
   (for example, of the Z boson at zero temperature), which is defined
   so that the probability (or equivalently the intensity or particle number)
   associated with the resonance decays as $\exp(-\Gamma t)$.  The
   relation is simply $\Gamma = 2 \gamma$.
}
Here, $\CA$ is the adjoint Casimir, conventionally normalized as $\CA = N$
for the gauge group SU$(N)$,
and ``$\approx$'' means equality up to relative corrections suppressed
by powers of $\ln(1/g)$.  That is, no claim is made about discriminating
$\ln(1/g)$ from $\ln(2/g)$.%
\footnote{
   It is not clear whether the color conductivity even has meaning
   except as an approximate concept valid at the level of leading
   logarithms.  We do not know, for instance, of any directly measurable
   (gauge-invariant, non-perturbative) definition of the color conductivity.
}
The only place where the matter content of the theory enters is in the
value of the plasma frequency.  For hot electroweak theory with a single
Higgs doublet, it is given by
\begin {equation}
   m_\pl^2 = {(5 + 2 \nf)\over18} \, g^2 T^2 \> [1+O(g)] \,,
\end {equation}
where $\nf$ is the number of fermion families.
The Langevin equation (\ref{eq:eff}) with the value
(\ref{eq:sigma}) of $\sigma$ precisely reproduces the effective theory
derived by B\"odeker \cite{Bodeker}.

It's interesting to note that, if the time is rescaled,
the Langevin eq.~(\ref{eq:eff}) is equivalent
to the stochastic quantization of three-dimensional Euclidean gauge
theory.%
\footnote{
   See, for example, chapter~17 of ref.~\cite{ZinnJustin}.
}
In that context, the time $t$ is usually considered a fictitious
additional variable, corresponding in simulations to Monte Carlo time.
Amusingly, the present application provides an instance where
Monte Carlo time for gauge theories is actually real time, up to
a calculable rescaling.


\section {Color conductivity}

\subsection {Qualitative Description}
\label {sec:roughsig}

We now review why the color conductivity depends on coupling as in
(\ref{eq:roughsigma}), and show how earlier estimates
\cite{ASY1,Huet&Son,Arnold}
of the time scale
for non-perturbative processes as $t \sim 1/(g^4 T)$
correspond to ignoring collision effects.
Begin by considering the current response to an external electric
field in a {\it collisionless} ultra-relativistic plasma.  For simplicity
of notation, consider a QED plasma for the moment rather than a
non-Abelian one.  If the external field were static and homogeneous,
particles in the plasma with charge $g$ would respond to the field
by a change in momentum
\begin {equation}
   \Delta \p = g \E \, \Dt
\end {equation}
over a time $\Dt$.  For small deviations, the change in velocity of
a typical particle whose energy is order $T$ would then be
\begin {equation}
   \Delta \v 
   \sim {\Delta \p \over p^0}
   \sim {g \E \, \Dt \over T}
   ,
\label {eq:sigma1}
\end {equation}
and the resulting current would be
\begin {equation}
   \J \sim n \, g \, \Delta \v \sim (g^2 T^2 \Dt) \, \E \,,
\label{eq:J1}
\end {equation}
where $n \sim T^3$ is the density of hard particles.
The current is dominated by the most prevalent particles in the plasma:
those with momentum of order $T$.
The current (\ref{eq:J1})
grows indefinitely with the length of time the electric field is applied.
There are two things which can cut off this growth of
the current: (a) collisions, and (b) temporal or spatial oscillation of
the electric field.  Stick with the collisionless plasma for a moment
and consider oscillations of $\E$.  As we've discussed, the time scale for
non-perturbative processes turns out to be slow.
So suppose, for example, that the electric field varies in
the $z$ direction as $\E \sim \E_0 \cos(kz)$ but not significantly in time.
Then current carriers, which have an rms $z$ velocity of $1/\sqrt{3}$,
will move from regions of positive $E_z$ to
regions of negative $E_z$ in a time of order
\begin {equation}
    \Dt \sim k^{-1} .
\end {equation}
This change in direction of the electric field felt by the charge carriers then
limits the average current response to a magnitude
\begin {equation}
J \sim {g^2 T^2 \over k} \, E .
\end {equation}
If we identify the ($k$-dependent) conductivity as
\begin {equation}
   \qquad\qquad\qquad\qquad\qquad\qquad
   \sigma(k) \sim {g^2 T^2 \over k}
   ,
   \qquad\qquad\qquad \hbox{(collisionless)}
\label {eq:sigmak}
\end {equation}
and take $k$ to be
the inverse spatial scale $g^2 T$ for non-perturbative physics,
then the time scale $t$ for non-perturbative physics would be
\begin {equation}
   t \sim {1 \over g^2 k} \sim {1\over g^4 T} \,,
\end {equation}
provided we could indeed ignore the effects of collisions on the conductivity.
This is the qualitative physics behind the more formal and quantitative
discussions of refs.\ \cite{ASY1,Huet&Son,Arnold}.%
\footnote{
   $\sigma(k)$ corresponds exactly to the damping coefficient
   $\gamma$ introduced in ref.\ \cite{Arnold}.
}

The divergence of the conductivity (\ref{eq:sigmak}) as $k \to 0$ is cut off
in real physical systems by the effects of collisions, as pointed out by Drude
in 1900.  Let's continue to focus on a QED plasma for the moment.  A charge
accelerated by the electric field eventually experiences a collision with
other particles in the plasma which changes the charged particle's direction
by a large angle.  Such collisions randomize the direction of the
particle and so randomize its contribution to the current.
So the relevant time
$\Dt$ determining the conductivity (\ref{eq:sigma1}) becomes the
mean collision time $\taularge$ for large angle scatterings:
\begin {equation}
    \taularge^{-1} \sim g^4 T \ln\left(T \over m_\pl\right) .
\end {equation}
The $g^4$ above just comes from the square of the scattering matrix element.
The logarithm arises because the randomization of the velocity can
occur either through a single large-angle scattering or through the
cumulative effect of many (individually more probable)
small-angle scatterings.%
\footnote{
  For a slightly more detailed but still qualitative summary see, for
  example, section III of ref.\ \cite{jj}.
  $\taularge$ is also known as the ``momentum relaxation'' time
  (see, for example, ref.\ \cite{taularge}).
}
If $\taularge$ were the relevant mean free time in the non-Abelian case,
then the effects of collisions on the conductivity $\sigma(k)$ could
safely be ignored when investigating non-perturbative fluctuations.
That's because $\taularge \gg 1/k \sim 1/g^2 T$, and so it
would be the collisionless time scale $1/k$ instead of
$\taularge$ that determines $\Dt$ and hence $\sigma(k)$.

\begin {figure}
\vbox
   {%
   \begin {center}
      \leavevmode
      
      \epsfbox {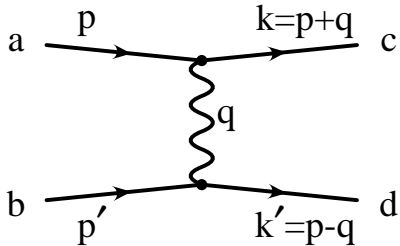}
   \end {center}
   \caption
       {%
          The dominant scattering process:
          $t$-channel gauge boson exchange.  The solid lines represent any
          sort of hard particles, including gauge bosons themselves.
          The labels $a,b,c,d$ show our convention for naming color indices
          of the various lines.
       \label{fig:tchannel}
       }%
   }%
\end {figure}

However, Selikhov and Gyulassy \cite{Selikhov} have pointed out that
$\taularge$ is not the relevant mean free time in the non-Abelian case.
In the non-Abelian case, even arbitrarily small angle scatterings can
randomize the current, not by randomizing the velocity of the current
carriers, but by randomizing their color charge.  The crucial difference
with QED is that an exchanged non-Abelian gauge boson, no matter how
soft, carries color and so changes the color charge of the scatterers,
whereas an exchanged photon is neutral.  The relevant
time scale for the non-Abelian case
is then the mean free time $\tauany$ for {\it any}-angle scattering,
which is much shorter than the mean free time for large-angle scattering.
Specifically,
$t$-channel gauge boson exchange, shown in fig.\ \ref{fig:tchannel},
gives a cross-section $\hat\sigma$ such that
\begin {equation}
  \tauany^{-1} \sim n\hat\sigma \sim n g^4 \int {d \tM \over \tM^2} \,,
\label{eq:tauany1}
\end {equation}
where $n \sim T^3$ is the density of particles and
$\tM = -Q^2$ is the virtuality of the exchanged gauge boson.
$\tauany^{-1}$ is also known as the thermal damping rate of the hard
particle carrying the current \cite{gammag,smilga,BI2}.
For $\tM$ below $m_\pl^2 \sim (g T)^2$, screening effects in the plasma
turn out to reduce the linear $\tM \to 0$ divergence in
(\ref{eq:tauany1}) to a logarithmic one.
The result is then that%
\footnote{
   Again, for more qualitative detail, see the review in section III of
   ref.\ \cite{jj}.  For the original work, see ref.~\cite{gammag}.
}
\begin {equation}
   \tauany^{-1} \sim {n g^4\over  m_\pl^2} \ln\left(m_\pl \over g^2 T\right)
                \sim g^2 T \ln\left(1 \over g\right) ,
\end {equation}
where the scale $k \sim g^2 T$ of non-perturbative physics has been used as
an infrared cut-off.
Using (\ref{eq:sigma1}) and comparing $\tauany$ to the collisionless time
scale $1/k$,
the zero-frequency conductivity $\sigma \sim g^2 T^2 \Dt$ is then
\begin {equation}
   \sigma(k) \sim \cases{ g^2 T / k ,       &   $k \gtrsim \tauany^{-1}$;\cr
                          g^2 T \, \tauany , &  $k \lesssim \tauany^{-1}$.}
\end {equation}
$\tauany$ wins by a logarithm for $k \sim g^2 T$.
This means that, in the small coupling, large logarithm limit, the
$k \to 0$ value of the conductivity,
namely $\sigma \sim T/\ln(1/g)$,
is what is relevant to non-perturbative
physics in non-Abelian plasmas.

\begin {figure}
\vbox
   {%
   \begin {center}
      \leavevmode
      
      \epsfbox {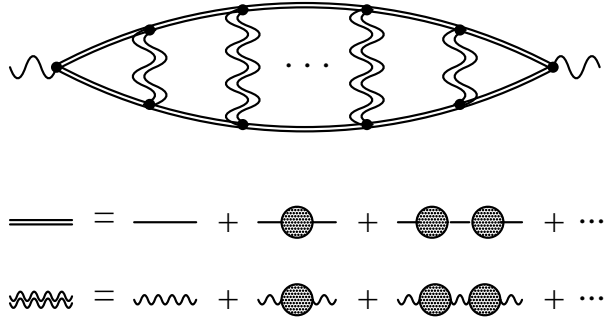}
   \end {center}
   \caption
       {%
          The Feynman diagrams (assuming finite-temperature Feynman rules)
          that produce the conductivity due to hard
          excitations at leading-log order.  Specifically, the ladder diagrams
          are for the self-energy of the soft fields, whose imaginary part
          is proportional to
          the conductivity at low frequency (it's $\omega\sigma$ in
          $A_0 = 0$ gauge).
          The external lines have soft momentum ($g^2 T$)
          and softer frequencies, the solid lines correspond to any type
          of colored particle with hard momentum ($T$), and the rungs have
          semi-hard momentum ($g^2 T\ln g^{-1} \ll q \ll g T$).
	  The double lines
          indicate that the dominant one-loop contributions to the
          self-energies have been included in the propagators.
          The analog of the two-loop chain diagram of $\phi^3$ theory
          \protect\cite{Jeon,Jeon&Yaffe}
          is not included because we only integrate out hard and semi-hard,
          but not soft, fields to obtain the effective theory of interest.
          Other diagrams relevant to $\phi^3$ theory ({\it e.g.}, non-pinching
          boxes and chain diagrams) have been dropped because they do not
          correspond to $t$-channel scattering and so should be sub-leading
          in the gauge-theory case.
       \label{fig:diagram}
       }%
   }%
\end {figure}

Some readers may want to know what Feynman diagrams, in the underlying,
fundamental quantum field theory, correspond to the color conductivity
discussed above.  In the next section, we formulate a leading-log
calculation of the
conductivity in terms of the Boltzmann equation.  Based on (a) the analogy of
QCD with scalar $\phi^3 + \phi^4$ theory (both have 3- and 4-point
interactions), (b) the diagrammatic analysis of transport coefficients for the
latter theory and its equivalence to the Boltzmann equation as explained
in refs.\
\cite{Jeon,Jeon&Yaffe}, and (c) the fact that, in the gauge theory case,
only $t$-channel
scattering processes are relevant at the order of interest, we believe
that the
relevant series of Feynman diagrams are the ladder diagrams
shown in fig.~\ref{fig:diagram}.
This is similar to the class of diagrams considered in
ref.\ \cite{smilga} for QED.
Diagrammatic perturbation theory in this form is awkward and cumbersome,
however, and we shall avoid it.


\subsection {Quantitative Description}

The original calculation of the color conductivity
by Selikhov and Gyulassy \cite{Selikhov} was clever but not absolutely
convincing.  For one thing, it was based on an approximation to the
evolution of color distribution functions which assumes that there is no
coupling between the different velocity components of a fluctuation.
(We shall explain more clearly what this means below.)  The approximation
is incorrect in general but, as we shall see, does not affect the
calculation of the color conductivity in particular.
Subsequently, Heiselberg \cite{Heiselberg} analyzed the quark
contribution to the conductivity by starting with a Boltzmann equation
with an appropriate collision term.  He obtained the same dependence on
coupling $g$ as Selikhov and Gyulassy but a different numerical coefficient.
As we shall later explain, this difference was primarily due to the
use of a plausible but inadequate variational ansatz.
Selikhov and Gyulassy \cite{Selikhov2a}
subsequently published an alternative
derivation of the color conductivity that also started from the Boltzmann
equation with a collision term \cite{Selikhov2b}.
Unfortunately, the collision term they used
did not account for quantum statistics of the hard
particles and, as a result, they were unable to obtain a final
answer without making some very rough approximations along the way.%
\footnote{
   Such as the approximation made in and just above eq.~(15) of
   ref.\ \cite{Selikhov2a}.
}
Indirectly, therefore, B\"odeker's \cite{Bodeker}
results seem to be the first complete (albeit terse)
analysis of the color conductivity, even though that is not the language he
uses.  We present here our own direct derivation of
the conductivity, based on the collision term approach, which
will be more familiar to some readers (and so perhaps more comforting)
than B\"odeker's methods.

There are three scales relevant to understanding the conductivity at
leading-log order.  Following B\"odeker, we will label them as (a)
the hard scale, corresponding to momentum $T$, and characteristic of
the charges which carry the current, (b) the soft scale, corresponding
to momentum $g^2 T$, characteristic of the non-perturbative electric
fields that the hard charge carriers respond to, and (c) the
{\it semi-hard} scale, corresponding to momenta $q$ in the range
$g^2 T \ll q \lesssim gT$, which (at leading-log order)
is the momentum scale of the
$t$-channel gauge bosons that mediate color randomization of the hard
charge carriers.  Remember that the logarithm in the conductivity is
a logarithm of the plasma frequency scale $gT$
over the soft scale $g^2 T$.
If we tried to go beyond leading-log order, then
the distinction between semi-hard and soft would blur, because
soft gauge bosons can also mediate color-randomizing processes.

We will describe the hard, perturbative modes of the
theory by a Boltzmann equation.  This is well known to reproduce exactly
(at leading order in coupling) a large variety of thermal results obtained
by a more fundamental analysis of diagrammatic perturbation theory.%
\footnote{
   For an explicit discussion of the relationship between diagrammatic
   perturbation theory and the Boltzmann equation in scalar theory,
   see ref.~\cite{Jeon&Yaffe}.
}
This kinetic description is valid whenever the mean free
time is long enough that the hard particles can be
treated as propagating classically ({\it i.e.}, on-shell)
between collisions---a condition to be discussed momentarily.
We will couple the hard particles to a soft electric field, and we will
incorporate the semi-hard mediated
scattering processes into the collision term.

The requirement that particles propagate classically between collisions
means that the de Broglie wavelength and the collision times
must be small
compared to the mean free path and mean free time, respectively.
The relevant mean free time
(and path) here is $\tauany \sim (g^2 T \ln)^{-1}$,
where here, and henceforth, we use ``ln'' as shorthand for $\ln g^{-1}$.
The de Broglie
wavelength of the hard particles is order $1/T \ll \tauany$.
The duration of collisions%
\footnote{
   One way to estimate the scattering duration is to consider specific
   time orderings of
   fig.\ \ref{fig:tchannel} and to estimate the energy difference $\Delta E$
   between the initial and intermediate states.  The duration is
   then order $1/\Delta E$.  Alternatively, one may consider a Feynman
   diagram representing two successive collisions of a particle and
   verify the requirement that a typical particle will be sufficiently
   on-shell between collisions that no error (at the desired order)
   is made by treating the collisions separately.
}
mediated by semi-hard gauge bosons is order
$1/|q \pm q_0| \sim 1/q$ and is small compared to $\tauany$
when $g^2 T\ln \ll q$.
This requirement means that we can only
properly account for scatterings with semi-hard momentum transfer
$q$ having $g^2 T \ln \ll q \lesssim gT$ rather than $g^2 T \ll q \lesssim gT$.
This will not, however, affect results at leading-log order,
which does not distinguish between $\ln(g T / g^2 T)$ and
$\ln(g T / g^2 T \ln)$.


\subsubsection{The Boltzmann (Waldmann-Snider) equation}

To introduce the Boltzmann equation we will use, let's start by ignoring
the details of color and the non-Abelian nature of the problem.  Pretend,
for the moment, that we were interested in the Boltzmann equation for
hard particles in a QED plasma.
Schematically, the Boltzmann equation for the local distribution
$n(\x,\p,t)$ of hard particles is of the form
\begin {equation}
   {d n \over dt}
   = - C[n] ,
\end {equation}
where $C[n]$ is a collision term describing the net loss,
due to semi-hard scattering, of hard particles with momentum $\p$.
The total time derivative can be rewritten in terms of a convective
derivative and the force exerted by the soft fields as
\begin {equation}
   {d n \over dt} \equiv
   (\partial_t + \dot\x \cdot \grad_\x + \dot\p \cdot\grad_\p) \, n =
   \partial_t n + \v\cdot\grad_\x \, n + g(\E + \v\times\B)\cdot\grad_\p \, n ,
\label{eq:dndt}
\end {equation}
where $\E$ and $\B$ are to be understood as soft.
[The difference between this and B\"odeker's approach \cite{Bodeker}
is that B\"odeker starts with a collisionless Boltzmann equation
($dn/dt = 0$) but includes coupling to dynamical semi-hard fields in
(\ref{eq:dndt}).]

The collision term is dominated by $2\to2$ collisions
and has the form%
\footnote{
   Our notation is $\int_\p \equiv \int{d^3p\over(2\pi)^3}$ for momentum
   integrals and $\int_\x \equiv \int d^3x$ for position integrals.
   With this convention, transition matrix elements $M$ 
   should be understood to have non-relativistic rather than
   covariant normalization.
}
\begin {mathletters}
\begin {eqnarray}
   C[n] &=& \int_{\p'\k\k'} |M_{\p \p' \to \k\k'}|^2
       \left[
             n_\p n_{\p'} (1 \pm n_\k) (1 \pm n_{\k'}) -
             n_\k n_{\k'} (1 \pm n_\p) (1 \pm n_{\p'})
	\right]
\\
        &\equiv& n_\p \, \I_- - (1 \pm n_\p) \, \I_+ ,
\end {eqnarray}%
\label{eq:C1}%
\end {mathletters}
where $M_{\p \p' \to \k\k'}$ is the matrix element for the collision,
and the $1{\pm}n$ are final-state Bose enhancement or Fermi blocking
factors, depending on whether the hard particles are bosons ($+$) or
fermions ($-$).
The first term of $C[n]$ in (\ref{eq:C1}) is a loss
term, representing scattering out of momentum state $\p$, and the second
term is a gain term, representing scattering into state $\p$.
Coefficients of the loss and gain terms,
$\I_\mp$, have been introduced for later convenience.
In equilibrium, $C[n]$ vanishes.

We have not included the coupling of
the soft electromagnetic fields to the spin
of the hard particles in (\ref{eq:dndt}).  There are a number of
independent reasons for
this: (a) the small-angle scatterings that determine the conductivity
are insensitive to the spins of the colliding particles, (b) such terms
vanish when one linearizes the Boltzmann equation \cite{Blaizot&Iancu},
as we shall eventually
do, and (c) for hard massless quarks, at least, the spin dynamics is made
trivial by conservation of helicity.
See ref.~\cite{deGroot,Blaizot&Iancu}\ for a discussion of including
spin effects in the Boltzmann equation.

We must now face the one subtlety in this derivation, which is how to
incorporate color into the collision term.
It's easy to put flavor indices into a collision term if all
distribution functions $n$ are diagonal in flavor: one must simply
use the specific matrix elements for flavors $a,b$ to collide and produce
flavors $c,d$ and then sum appropriately over flavor indices.
The problem is more subtle for color, however, since,
as we will review, the distribution functions that describe
color fluctuations are {\it not} diagonal in color space.
The need to deal with a non-diagonal distribution $n$ is a problem
that has arisen previously in applications involving
massive particles with spin: if one quantizes spin in
the $z$ direction, there is then no way to describe
spin 1/2 particles with spin, say, in the $x$ direction in terms of definite
numbers $n_{\pm}$ of particles with spin in the $\pm z$ direction.
In fact, the problem goes back to the physics of dilute
gases of molecules with spins that, between collisions, precess in
an external magnetic field; the generalization of the Boltzmann equation
which solved that problem is known as the Waldmann-Snider equation.%
\footnote{
   For a review, see ref.\ \cite{Snider}.
}
We need the appropriate generalization to the
problem at hand.

In preparation, let's
review
the incorporation of color
into the collisionless part (\ref{eq:dndt}) of the Boltzmann equation,
which is well known \cite{adjXadj}.
We will give a quick summary, rather than starting from
first principles.  The first thing to note is that number operators
for particles are of of form $a_\p^\dagger a_\p$ in terms of creation
and annihilation operators.%
\footnote{
   More technically, the localized number densities $n(\x,\p)$
   correspond to expectations of the Wigner operators
   $a_{\p+\k/2}^\dagger a_{\p-\k/2}$ where $\k$
   is the Fourier transform variable conjugate to $\x$
   and should be regarded as small compared to $\p$.
}
Since both $a$ and $a^\dagger$ carry color indices, we see that $n$
is a matrix and
transforms under color as $\R \times \bar\R$ if the hard particles
are in the representation $\R$.  To generalize the convective
derivative in (\ref{eq:dndt}) to the non-Abelian case, gauge-invariance
then requires the derivatives $\partial_t$ and $\grad_\x$ to be
replaced by gauge-covariant derivatives $\cD_t$ and $\cD_\x$ acting
in the $R \times \bar R$ representation.  That is,
\begin {equation}
   \partial_\mu n \to \cD_\mu n = \partial_\mu n - ig \, [A_\mu, n] ,
\end {equation}
where $A_\mu$ is the soft gauge field expressed in terms of generators of the
representation $\R$ of the hard particles.  The third term in
(\ref{eq:dndt})---the electromagnetic force term---could be color
contracted as either $[E,\nabla_\p n]$ or $\{E,\nabla_\p n\}$, with
$E$ expressed in terms of generators of $\R$.  The fact that $n$, and
so also $dn/dt$, are Hermitian
rules out the commutator.  So the non-Abelian Boltzmann equation is
\begin {equation}
   (\cD_t + \v\cdot\cD_\x) \, n
   + \half g\,\{(\E + \v\times \B)_i, \grad_{p_i} n\}
   = -C[n] ,
\label{eq:boltz}
\end {equation}
where we have yet to specify the collision term $C[n]$.
In Appendix \ref{app:collision},
we discuss how to generalize the collision term (\ref{eq:C1}).
(The result is substantively equivalent to one derived by Botermans
and Malfliet \cite{Botermans}
in the context of one-boson exchange processes in nuclear
matter.%
\footnote{
   Botermans and Malfliet, however, absorb the ${\rm Re} \Sigma$ term in
   (\ref{eq:C}) by redefining their flavor states to diagonalize the
   effective Hamiltonian.
}%
)
Here, we will just try to make the result plausible.  It is fairly easy
to guess how to contract all the color indices in (\ref{eq:C1}) {\it other\/}
than the one whose net loss is being described (that is, other than
the one associated with $\p$).  Refer to fig.\ \ref{fig:tchannel}:%
\footnote{
   We make no particular distinction between upper and lower color indices.
}
\begin {eqnarray}
   \I_-^{a \bar a} &=&
            \int_{\p'\k\k'} M^*_{\bar a\bar b\bar c\bar d} M_{abcd} \>
            n_{\p'}^{\bar b b} \,
	    (1 \pm n_\k)^{c \bar c} \, (1 \pm n_{k'})^{d \bar d} \,,
   \label{eq:I-}
\\
   \I_+^{a \bar a} &=&
            \int_{\p'\k\k'} M^*_{\bar a\bar b\bar c\bar d} M_{abcd} \>
            n_\k^{c \bar c} \, n_{\k'}^{d \bar d} \,
	    (1 \pm n_{\p'})^{\bar b b} \,.
   \label{eq:I+}
\end {eqnarray}
Here and henceforth, there will always be an implied summation over the types
and spins of the particles associated with $\p'$
(quarks, anti-quarks, gauge bosons, Higgs, etc.).
In terms of $\I_\pm$, the correct collision term then turns out to be
\begin {equation}
   C[n] = \half \{n_\p,\I_-\} - \half \{1\pm n_\p,\I_+\}
   - i \, [{\rm Re} \, \bar\Sigma, n_\p] ,
\label{eq:C}
\end {equation}
where all commutators are in color space and $\bar\Sigma$ is the self-energy
of the hard particles (non-relativistically normalized).

The appearance of the self-energy term is easy to
understand, although it will disappear when we linearize the Boltzmann
equation.
Time evolution of observables,
ignoring dissipation, is given by
\begin {equation}
   {dA \over dt} = i \, [H_\eff,A] ,
\end {equation}
and the the real part of the self-energy contributes
to the effective Hamiltonian.
The loss and gain terms are related to the imaginary
part of the self-energy, and a simple mnemonic
(although hardly a real derivation)
for the appearance of anti-commutators in those terms is to consider the
time-evolution of an observable with a non-Hermitian effective Hamiltonian:
\begin {equation}
   A(t) = U^\dagger(t) \, A(0) \, U,
   \qquad\qquad U(t) = e^{-i H_\eff t},
   \qquad\hbox {and}\qquad H_\eff = R + i I ,
\end {equation}
so that
\begin {equation}
   {dA \over dt} = \{I,A\} + i \, [R,A] .
\end {equation}
This is the same sort of structure that appears in the collision term.

If the final-state statistical factors are ignored, so that
$1\pm n \to 1$, then (\ref{eq:C}) has exactly the form of the
relativistic collision term presented in ref.~\cite{deGroot}%
\footnote{
   Specifically, eq.\ (26) of section B.IV.3 of ref.\ \cite{deGroot}.
}
for spin (as opposed to color) degrees of freedom.


\subsubsection{The linearized Boltzmann equation}

We will now linearize the Boltzmann equation, since the fluctuations
in the hard particles induced by soft fields are small (as parameterized
by powers of the coupling).  Write
\begin {equation}
   n^{a \bar a} = n_\eq \, \delta^{a \bar a} + \delta n^{a \bar a} ,
\end {equation}
where $n_\eq$ is the equilibrium distribution and is colorless.
The linearization of the Boltzmann equation given by (\ref{eq:boltz})
is
\begin {equation}
   (\cD_t + \v\cdot\cD_\x) \, \delta n + g \, \E\cdot\v \, {dn_\eq\over dp}
   = -\delta C[\delta n] ,
\label{eq:boltzlin}
\end {equation}
where $\delta C$ is the linearization of (\ref{eq:C}).
The equilibrium self-energy must be colorless
(proportional to $\delta^{a \bar a})$,
and so the linearization
of the self-energy term in $\delta C$ vanishes:
\begin {equation}
   \delta[{\rm Re}\Sigma,n] =
   [\delta({\rm Re} \Sigma), n_\eq] + [{\rm Re} \, \Sigma_\eq, \delta n] = 0 .
\end {equation}
The linearization of
the loss and gain ($\I_\pm$) pieces of the collision integral
may be simplified by recalling that small-angle
collisions will dominate the physics.
The dominant momentum transfer $q$ lies between
$g^2 T$ and $g T$, and is small compared to the momenta of the colliding
hard particles.  So, to leading order in coupling, we can replaces
$n_\k = n_{\p+\q}$ and $n_{\k'} = n_{\p'-\q}$ by $n_\p$ and $n_{\p'}$.
The result of linearizing the collision term (\ref{eq:C}) in this
small momentum-transfer approximation is then%
\footnote{
   The collision term in (\ref{eq:Clin})
   is the same as the $\Delta C_2$ given by
   Selikhov and Gyulassy
   in eq.~(6) of ref.\ \cite{Selikhov2a} (originally derived by Selikhov
   \cite{Selikhov2b}) except for the statistical factors of $1 \pm n$.
}${}^,$%
\footnote{
   We have swept under the rug the fact that the matrix element depends on the
   self-energy $\Pi$ of the exchanged gauge boson, which in turn depends on
   the distribution functions $n$.  One should consider fluctuations of
   these distribution functions as well, but, at linear order in $\delta n$,
   these variations
   do not contribute to $\delta C$ because the loss and gain terms cancel.
}
\begin {equation}
   \delta C[\delta n] = \half \int_{\p'\q} |{\cal M}|^2 \,
   \Bigl\{
     \TRp [T_\R^a, [T_\R^a, \delta n_\p] ] \, n_{\p'} (1 {\pm} n_{\p'})
     - \CA T_\R^c \, \tr(T_{\R'}^c \, \delta n_{\p'}) \, n_\p (1 {\pm} n_\p)
   \Bigr\} ,
\label{eq:Clin}
\end {equation}
where here (and henceforth) we have dropped the subscript
``eq'' from the equilibrium distribution $n_\eq$.
The matrices $\{ T_\R^a \}$ are color generators for the representation $\R$;
${\cal M}$ is the $t$-channel matrix element of fig.\ \ref{fig:tchannel}
stripped of color generators, $M_{abcd} = {\cal M} \> T_\R^{ac} T_{\R'}^{bd}$;
and $\TR$ is the normalization constant defined by
\begin {equation}
   \tr(T_\R^a T_\R^b) = \TR \, \delta^{ab} ,
\end {equation}
which is $\CA$ for the adjoint representation and,
with conventional normalization,
$\half$ for the fundamental.

Note that the expression (\ref{eq:Clin}) for $\delta C$ vanishes for color
neutral fluctuations, {\it i.e.}, when $\delta n$ is proportional to the
identity.  To treat such fluctuations, one must expand $n_\k = n_{\p+\q}$
and $n_{\k'} = n_{\p-\q}$ to higher order in $\q$ than we have done,
which leads to suppression by more powers of $g$. (An example is
the difference between the inverse momentum relaxation time $\taularge^{-1}$
and the color relaxation time $\tauany^{-1}$ discussed earlier.)

For comparison to $\delta C$ (\ref{eq:Clin}), note that
the hard thermal damping rate defined
from the imaginary part of the self-energy in equilibrium is, to leading
order in coupling,\footnote{
   The overall factor of 1/2 arises because the damping rate is defined
   in the literature as
   the decay rate for the quantum-mechanical amplitude of an excitation
   rather than the decay rate for the number density of an excitation.
   See footnote \ref{fn:width}.
} 
\begin {eqnarray}
   \gamma_\R &=&
   \half \, {d \over d n_\p}
   \left\{ \int_{\p'\q} |{\cal M}|^2 \, C_\R \, \TRp \,
     [ n_\p \, n_{\p'} \, (1 {\pm} n_\k) \, (1 {\pm} n_{\k'})
     - n_\k \, n_{\k'} \, (1 {\pm} n_{\p}) \, (1 {\pm} n_{\p'})] \right\}
\nonumber\\ &=&
   \half\int_{\p'\q} |{\cal M}|^2 \, C_\R \, \TRp \,
     [ n_{\p'} \, (1 {\pm} n_\k) \, (1 {\pm} n_{\k'})
     \mp n_\k \, n_{\k'} \, (1 {\pm} n_{\p'})]
\nonumber\\
     &\simeq& \half\int_{\p'\q} |{\cal M}|^2 \, C_\R \, \TRp \,
     n_{\p'} \, (1 {\pm} n_{\p'}) ,
\label{eq:gammaR}
\end {eqnarray}
where in the last step we've used the small momentum-transfer limit (valid
at leading order in coupling).
As always, there is an
implicit summation over the particle type and spin
associated with $\p'$ in (\ref{eq:gammaR}).
Note that the coefficient of $\delta n$ in the first term of
eq.\ (\ref{eq:Clin}) for
$\delta C$
is, up to color factors, just the thermal damping rate $\gamma_\R$.

The representation $\R\times\bar\R$ we have been ascribing to fluctuations
$\delta n$ is reducible and a bit over-general for our needs.  There
is only one irreducible component of $\R\times\bar\R$ which contributes to
the conductivity---the adjoint representation.  There are a number of
ways to see this.  First, the color current $\J$ is given by
\begin {equation}
    \J^a = g \int_\p \tr(T_\R^a \, \delta n_\p) \, \v_\p
\end {equation}
(with implicit summation over particle type and spin),
and $\J$ only receives contributions from the pieces of $\delta n$ proportional
to the generators $T^a$.  Alternatively, on the left-hand side of the Boltzmann
equation (\ref{eq:boltzlin}), the driving term $\E\cdot\v\, (dn/dp)$ is in the
adjoint representation.  We may thus specialize $\delta n$
to fluctuations of the form
\begin {equation}
   \delta n_\R = T_\R^a \, \delta N^a .
\label{eq:nform}
\end {equation}
The linearized Boltzmann equation given by (\ref{eq:boltzlin}) and
(\ref{eq:Clin}) then becomes
\begin {equation}
   \left[(D_t + \v\cdot\D_\x) \, \delta N\right]^a
   + g\,\E^a\cdot\v \, {dn\over dp}
   =
   \half \, \CA \, \TRp \int_{\p'\q} |{\cal M}|^2
   \left\{
     \delta N_\p^a \, n_{\p'} (1{\pm}n_{\p'})
     - \delta N_{\p'}^a \, n_\p (1{\pm}n_\p)
   \right\} ,
\label{eq:boltzlin2}
\end {equation}
where the covariant derivatives $D_\mu$ now act in the adjoint representation.
Comparison with (\ref{eq:gammaR}) shows that the first term above is
just $\gamma_\g \, \delta N_\p^a$, where $\gamma_\g$ is the
thermal damping rate of hard gauge bosons.
The color current resulting from $\delta N^a$ is
\begin {equation}
   \J^a = g \, \TR \int_\p \delta N^a \, \v .
\end {equation}


Now let us finally turn to the matrix element ${\cal M}$.
At small momentum transfers, the classic Coulomb scattering amplitude
may be written in the form
\begin {equation}
   \int_\q |{\cal M}|^2
    = g^4 \int {d^4 Q\over (2\pi)^4} \>
           {\left|V_\mu \, {\delta^{\mu\nu} \over Q^2} \, V'_\nu\right|^2}
           \> 2\pi \delta(Q\cdot V) \> 2\pi \delta(Q'\cdot V)
    ,
   \qquad \hbox{(no screening)}
\end {equation}
where $Q = (q_0,\q)$ and $V = (1,\v)$.
However, as discussed earlier, the conductivity is dominated by
momentum transfers $q$ small enough that plasma screening effects are
important.  In particular, the momentum range of relevance at leading-log
order is $g^2 T \ll q \ll g T$.  In this
regime, longitudinal forces are Debye screened, and hard particles only
interact through transverse (magnetic) forces.  The above $\delta^{\mu\nu}$
should therefore be replaced by the transverse projection operator
$\delta^{ij} - q^i q^j$, and the transverse self-energy $\PiT$ should
be resummed into the propagator of the exchanged gauge boson:
\begin {equation}
   \int_\q |{\cal M}|^2
    \approx g^4 \int {d^4 Q\over (2\pi)^4} \>
           {\left|v_i \,{(\delta_{ij} - q_i q_j) \over Q^2 + \PiT(Q)}\,
                  v_j\right|^2}
           \>2\pi \delta(q_0 - \q\cdot\v) \> 2\pi \delta(q_0 - \q\cdot\v')
    ,
\label{eq:Msqr1}
\end {equation}
where the $\approx$ sign indicates we've now made approximations valid
only at the leading-log level.  The full one-loop result for $\PiT(Q)$ is
well known \cite{Pi}, but we shall see in a moment that we need only its
$q_0 \ll \q$ limit.  In that domain, it is simply
\begin {equation}
   \PiT(Q) \approx i \sigma_0(q) \, \omega
\end {equation}
where
\begin {equation}
   \sigma_0(q) \equiv {3 \pi m_\pl^2\over 4 q}
\end {equation}
is the collisionless conductivity discussed earlier.  As also
discussed earlier, the collisionless approximation is in fact only valid
for $g^2 T \ln \ll q$ rather than $g^2 T \ll q$.  (Diagrammatically,
the breakdown for $g^2 T \lesssim q \lesssim g^2 T \ln$ appears
as a failure of the one-loop approximation to $\PiT$.)  So our
current approximations are really only valid for
$g^2 T \ln \ll q \ll g T$.  As we shall see shortly,
this will not affect the result for (\ref{eq:Msqr1}) at leading-log order.

Given these approximations, the $q$ integration in (\ref{eq:Msqr1}) is
dominated by $q_0 \sim q^2/\sigma_0(q) \ll q$, justifying the small $q_0$
approximation.  Performing the $q_0$ and angular integrations first,
one obtains
\begin {equation}
   \int_\q |{\cal M}|^2
    \approx {32 \alpha^2 \over 3 m_\pl^2} \>
            {(\v\cdot\v')^2 \over \sqrt{1-(\v\cdot\v')^2}}
            \int {d q \over q} .
\end {equation}
The logarithmic integral is cut off by $g T$ on one side (above
which the un-approximated integrand starts to fall more rapidly)
and the soft scale $g^2 T$ or the inverse collision time
$g^2 T \ln$ on the other side---it doesn't matter which.
The result at leading-log order is
\begin {equation}
   \int_\q |{\cal M}|^2
    \approx {32 \alpha^2 \over 3 m_\pl^2} \>
            \ln (g^{-1}) \>
            {(\v\cdot\v')^2 \over \sqrt{1-(\v\cdot\v')^2}} .
\label{eq:Msqr2}
\end {equation}

At this point, we have all the elements we need.  To proceed, it
is convenient to follow
B\"odeker \cite{Bodeker} and others \cite{W}
and combine
the different color distribution functions $\delta N^a$ for different
particles and different $|\p|$ by noticing that the current $\J$
depends only on the combination
\begin {equation}
   W^a(\x,\v,t) \equiv {g \over 3 m_\pl^2} \sum_\typespin
    \int {4\pi|\p|^2 d|\p| \over (2\pi)^3} \>
    \TR \, \delta N^a(\x,\p,t) ,
\end {equation}
where we have integrated over $|\p|$ but not $\v \equiv \hat\p$.
Integrate and sum both sides of the Boltzmann
eq.~(\ref{eq:boltzlin2}) similarly.
Then, making use of the value
\begin {equation}
   m_\pl^2 = {g^2 \over 3T} \sum_\typespin \int_\p \TR \, n_\p (1 \pm n_\p)
\end {equation}
and of the fact that the result (\ref{eq:Msqr2}) for $\int_\q |{\cal M}|^2$
depends only on angles, and comparing to the adjoint representation
expression $\gamma_\g$ of the hard thermal damping rate (\ref{eq:gammaR}),
one obtains
\begin {mathletters}%
\label{eq:boltzW}%
\begin {equation}
   (D_t + \v\cdot\D_\x) \, W - \E\cdot\v
   = -\delta C[W] ,
\label{eq:boltzWa}
\end {equation}%
\begin {equation}
   \delta C[W](\v) =
   \gamma_\g \left[
      W(\v)
      - {1\over\pi^2} \left\langle
            {(\v\cdot\v')^2 \over \sqrt{1-(\v\cdot\v')^2}} \,
            W(\v')
        \right\rangle_{\v'}
    \right] ,
\label{eq:dCW}
\end {equation}%
\end {mathletters}%
and
\begin {equation}
   \J = 3 m_\pl^2 \left\langle \v W(\v) \right\rangle_\v ,
\label{eq:JW}
\end {equation}
where $\langle \cdots \rangle_\v$ indicates angular averaging over $\v$.
This precisely reproduces the result that B\"odeker derived by another
method \cite{Bodeker}.

It is the second term in (\ref{eq:dCW}) that was dropped in the original
analysis of Selikhov and Gyulassy \cite{Selikhov}.  It is relevant
to some aspects of color dynamics, a prime example (noted by B\"odeker)
being the conservation $D_\mu J^\mu = 0$ required of the hard current
$J^\mu = 3 m_\pl^2 \langle V^\mu W(\v) \rangle_\v$ by the effective
Maxwell equation $D_\mu F^{\mu\nu} = J^\nu$ for the soft fields.
From (\ref{eq:boltzWa}), this conservation requires
$\langle \delta C[W] \rangle_\v = 0$, which is indeed satisfied by
(\ref{eq:dCW}).

The fact $\langle \delta C[W] \rangle_\v = 0$ can be rephrased to
say that the symmetric operator $\delta C$ has zero modes: it
annihilates anything that is independent of $\v$.
(
This can be rephrased in bra-ket notation
in $\v$-space as
$\langle \hbox{constant} | \delta C | W \rangle =
\langle W | \delta C | \hbox{constant} \rangle = 0$ for any $W$.)


\subsubsection{The conductivity}
\label {sec:sigresult}

To solve the linearized Boltzmann equation (\ref{eq:boltzW}) for
the $W$ at leading-log order,
B\"odeker \cite{Bodeker} argues that the covariant derivative terms are
together
order $g^2 T \, W$ and so can be ignored compared to the collision term,
which is order $\gamma_\g W \sim (g^2 T \ln g^{-1})\,W$.
This approximation is actually flawed because of the zero mode of
$\delta C$.  We analyze this flaw in the approximation in
ref.\ \cite{paper2} and show that it does not affect the transverse
dynamics.  Here, we shall instead simply
continue with the naive approximation.
Dropping the covariant derivative terms from the
Boltzmann equation gives simply
\begin {equation}
   \E\cdot\v
   \approx \delta C[W] .
\label{eq:dropD}
\end {equation}%
Next note that $\delta C$ maps even (odd) functions
of $\v$ into even (odd) functions of $\v$.
(In contrast, the $\v\cdot\D_\x$ operator that we dropped does not.)
Since $\E\cdot\v$ is odd
in $\v$, the solution $W$ to (\ref{eq:dropD}) must be odd as well.
But for odd functions of $\v$, the form (\ref{eq:dCW}) of $\delta C$ simplifies
to $\delta C[W] = \gamma_\g W$.%
\footnote{
  This is gratifyingly simpler than the leading-log collision terms one
  obtains for most transport phenomena, where $\delta C$ reduces to
  a linear differential operator \cite{HeiselbergViscosity}
  and the linearized Boltzmann equation must be
  solved either numerically or approximately.
}
The solution is then
\begin {equation}
   W \approx {\E\cdot\v \over \gamma_\g} ,
\end {equation}
which inserted into (\ref{eq:JW}) generates a current
\begin {equation}
   \J \approx {m_\pl^2 \over \gamma_\g} \, \E .
\label{eq:JE}
\end {equation}
This is Selikhov and Gyulassy's leading-log result (\ref{eq:sigma}) for the
conductivity.
[We show in ref.\ \cite{paper2} that a more careful analysis
of the Boltzmann equation (\ref{eq:boltzW}) reveals
that the $\E$ above should really be the transverse projection
of $\E$.]


\subsubsection{Variational methods}

We are now in a position to understand the problem with the estimate
of the color conductivity made by Heiselberg in ref.\ \cite{Heiselberg}.
Heiselberg uses a variational method to approximately solve the
Boltzmann equation.  The variational ansatz he uses is one that works
stunningly well for the diffusion of global or Abelian charges.
Using the imagery of QED, one imagines the linear response of the system
as a simple boost of equilibrium distributed positive charges moving in
one direction and of negative charges moving in the other,
with the boost velocities depending on the charges of the particles:
\begin {equation}
   n_i = {1 \over e^{\beta(\epsilon_\p - \u_i \cdot \p)} \mp 1}
   \simeq n_i^\eq - {dn\over d\epsilon_\p} \, \u_i \cdot \p ,
\label{eq:ansatz}
\end {equation}
where $i$ is a flavor index. 
Eq.~(\ref{eq:ansatz}) is Heiselberg's ansatz, with the velocity
$\u_i$ to be determined variationally.

When deriving the conductivity, we found it convenient to combine all
particles together and work with $W$ instead of $\delta n$.
If one instead follows through the argument of section \ref{sec:sigresult}
with the original Boltzmann equation (\ref{eq:boltzlin2}) for
$\delta n$, one finds that
\begin {equation}
   \delta n \propto {dn\over d\epsilon_p} \, {\E \cdot \v}
\label{eq:response}
\end {equation}
at leading-log order.
For color diffusion,
Heiselberg's ansatz (\ref{eq:ansatz}) misses the mark by a factor
of $|\p|$.  The actual linear response (\ref{eq:response})
cannot be interpreted as simple
boosts of fluids corresponding to different charges.


\section {Ultraviolet Insensitivity}
\label {sec:uv}

We will now elaborate on our earlier claim that B\"odeker's effective theory
(\ref{eq:eff}) of the soft modes has the wonderful property that it is
insensitive to how the soft modes are cut off in the ultraviolet.
Equivalently, but more technically, the effective theory does not
require any ultraviolet renormalization---it is ultraviolet finite.

As preparation, let us ignore the dynamics for a moment and remember that the
equilibrium properties of the classical theory are described by the
partition function
\begin {equation}
    {\cal Z} = \int [{\cal D}A] \> e^{-\beta \V}
\end {equation}
where
\begin {equation}
   \V = \half \int_\x B^2 = \fourth \int_\x F^a_{ij} F^a_{ij}
\end {equation}
is the potential energy associated with the magnetic field.
This is nothing other than the partition function for three-dimensional
Euclidean gauge theory, and three-dimensional gauge theory is ultraviolet
finite.  One way to see this is to rescale the
fields $\A$ and coupling $g$ (hiding in the definition of the field
strength) to absorb $\beta = T^{-1}$ by
\begin {equation}
   \A \to T^{1/2} \A ,
   \qquad\qquad
   g \to T^{-1/2} g .
\label {eq:scaleit}
\end {equation}
The partition function is then
\begin {equation}
    \int [{\cal D}A]\> \exp\left(-\fourth \int_\x F^a_{ij} F^a_{ij}\right) .
\label{eq:part}
\end {equation}
From (\ref{eq:scaleit}) or (\ref{eq:part}),
the gauge field can now be seen to have scaling dimension of
$[A] = \half$, and the coupling constant has dimension
$[g] = \half$.  There are no other relevant terms (in the sense of mass
dimension) that could be added to the action which are gauge
and parity invariant.  So the dimension-$\half$ coupling $g$ is the
{\it only} relevant parameter of this theory.
Now suppose we modify
or integrate out (in a gauge-invariant manner)
ultraviolet degrees of freedom associated with an
arbitrarily large momentum scale $\Lambda$.  We then potentially need
to modify (\ref{eq:part}) to
\begin {equation}
    \int [{\cal D}A]\> \exp\left(-\fourth Z \int_x F^a_{ij} F^a_{ij}\right) ,
\label{eq:partZ}
\end {equation}
where $Z$ is a renormalization constant.
However, it follows by dimensional analysis
that the perturbative expansion of $Z$ must be in powers of
$g^2/\Lambda$, which vanishes for $\Lambda \to \infty$.

Now let's turn to B\"odeker's effective theory (\ref{eq:eff}), which
we write in the generic form
\begin {mathletters}%
\begin {equation}
   \sigma {d\over dt} \, \A = - {\delta\over\delta\A} \, \V + \bnoise \,,
\end {equation}%
\begin {equation}
   \langle \noise(t,\x) \noise(t',\x') \rangle
         = 2 \sigma T \, \delta(t-t') \, \delta(\x-\x') \,,
\end {equation}%
\end{mathletters}%
suppressing color and vector indices.
By rescaling fields and coupling as before, and also rescaling time by
\begin {equation}
   t \to \sigma t ,
\end {equation}
one can put this in the form
\begin {mathletters}%
\begin {equation}
   {d\over dt} \, \A = - {\delta\over\delta\A} \V + \bnoise ,
\end {equation}%
\begin {equation}
   \left\langle \noise(t,\x) \noise(t',\x') \right\rangle
         = 2 \delta(t-t') \, \delta(\x-\x') .
\end {equation}%
\end{mathletters}%
Once again, the theory appears to depend on only one parameter:
the dimension-$\half$ coupling $g$.
The essential point to understand is that no other relevant terms can be
added to this equation---that is, more complicated time dependence
(which survives when the cutoff scale $\Lambda\to\infty$)
cannot be generated for the long-distance modes
when one modifies or integrates out short-distance physics.
This has been proven in a general analysis of the renormalizability
of purely dissipative stochastic field equations by Zinn-Justin and Zwanziger
\cite{ZZ}.%
\footnote{
  For a review, see chapters 16 and 17, and especially section 17.5.2,
  of ref.\ \cite{ZinnJustin}.
}
They analyze the problem by first finding a path-integral representation of
the stochastic equation, and then using dimensional analysis and various
BRS symmetries to determine the allowed relevant terms.  When translated
back into a stochastic equation, the result is that no more-complicated
time dependence can be generated and that a renormalized version of
B\"odeker's effective theory will take the form
\begin {mathletters}%
\begin {equation}
   Z_t \, {d\over dt} \A = - {\delta\over\delta\A} \, \VR + \bnoise ,
\end {equation}%
\begin {equation}
   \langle \noise(t,\x) \noise(t',\x') \rangle
         = 2 Z_t \, \delta(t-t') \, \delta(\x-\x') ,
\end {equation}%
\end{mathletters}%
where $\VR$ is the renormalized potential and $Z_t$ is a new renormalization
constant.
We already know that the potential is not renormalized.  And, just as
before, $Z_t$ must have a perturbative expansion in $g^2/\Lambda$
and so generates no relevant correction to the equation.  This demonstrates
why B\"odeker's equation is insensitive to the ultraviolet.


\section* {ACKNOWLEDGMENTS}

We are indebted to Dietrich B\"odeker, Henning Heiselberg, Guy Moore,
Christiaan van Weert, and Daniel Zwanziger for useful conversations.
This work was supported, in part, by the U.S. Department
of Energy under Grant Nos.~DE-FG03-96ER40956 and DF-FC02-94ER40818.


\appendix

\section {The collision term}
\label{app:collision}

In this Appendix we review the derivation of the Boltzmann equation from first
principles.
We will work up from the simplest case of a single-component $\phi^4$
theory, to multi-component scalar theories, and then to gauge theories.
We will discuss scalar QED first and finally derive the
Boltzmann equation for the case of primary interest,
non-Abelian gauge theory.

\subsection{Single-component $\phi^4$ theory}

Our starting point is the Schwinger--Keldysh closed-time-path (CTP) formalism
\cite{Schwinger,Keldysh}. Since both the Schwinger--Keldysh CTP formalism
and the derivation of the Boltzmann equation from it can be found in the
literature, the exposition here will be rather concise.  For more details,
see Ref.\ \cite{LifshitzPitaevskii}.

\begin{figure}[ht]
\setlength{\unitlength}{1cm}
\begin{center}
\begin{picture}(7,2)
\put(0,1){\line(1,0){7}}
\put(3,0){\line(0,1){2}}
\put(1,.9){\line(0,1){.2}}
\put(.9,.3){$t_0$}
\put(6.8,1.7){$t$}
\put(6.7,1.6){\line(1,0){.3}}
\put(6.7,1.6){\line(0,1){.4}}
\thicklines
\put(1,1.25){\vector(1,0){3}}
\put(4,1.25){\line(1,0){2}}
\put(6,0.74){\vector(-1,0){2}}
\put(4,0.74){\line(-1,0){3}}
\put(6,1){\oval(0.5,0.5)[r]}
\put(1.8,1.5){$C$}
\end{picture}
\end{center}
\caption{The Schwinger-Keldysh closed-time-path contour.}
\label{fig:app:contour}
\end{figure}
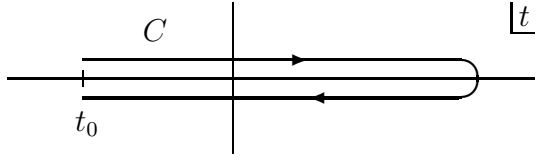

In the Schwinger--Keldysh formalism,
one considers time evolution along a time
contour, denoted $C$ in Fig.~\ref{fig:app:contour},
running from some initial time $t=t_0$ to $t=+\infty$ and
then returning to $t_0$.
Correspondingly, the propagator is defined as
$i\,G_C(x,y) = \langle {\cal T}_C (\phi(x)\phi(y))  \rangle$,
where ${\cal T}_C$ denotes contour ordering. Since both
$x$ and $y$ may lie on either the upper or the lower portion of the contour,
the propagator $G_C$ may be separated into 4 different components,
\begin{eqnarray}
  i\,G_{11}(x,y) &=& \langle {\cal T} (\phi(x)\phi(y)) \rangle, \qquad
  i\,G_{12}(x,y) = \langle \phi(y)\phi(x) \rangle, \nonumber \\
  i\,G_{21}(x,y) &=& \langle \phi(x)\phi(y) \rangle, \qquad\quad\;\;
  i\,G_{22}(x,y) = \langle \bar{{\cal T}} (\phi(x)\phi(y)) \rangle,
  \label{app:G-def}
\end{eqnarray}
where $\bar{{\cal T}}$ denotes anti-time-ordering.
From Eq.\ (\ref{app:G-def}) it is
apparent that the four components of $G_C$ satisfy the relation
\begin{equation}
  G_{11} + G_{22} = G_{12} + G_{21} \label{app:G-id} \,.
\end{equation}
It is also useful to introduce retarded and advanced propagators,
\begin{eqnarray}
  i\,G_R(x,y) & = &
  \theta(x_0{-}y_0) \left\langle [\phi(x), \phi(y)]\right\rangle ,
  \nonumber \\
  i\,G_A(x,y) & = &
  -\theta(y_0{-}x_0)\left\langle [\phi(x), \phi(y)]\right\rangle ,
\end{eqnarray}
which are related to the components of $G_C$ in the following ways,
\begin{eqnarray}
  G_R & = & G_{11}-G_{12} = G_{21}-G_{22} \,, \nonumber \\
  G_A & = & G_{12}-G_{22} = G_{11}-G_{21} \,. \label{app:GRA-def}
\end{eqnarray}
The retarded and advanced propagators are the boundary values
of the Euclidean time-ordered propagator when the imaginary
(Matsubara) frequency $i\omega_n$ is analytically continued to just above,
or just below, the real frequency axis.

For a free scalar field with Lagrangian $L = \half (\partial_\mu\phi)^2 -
\half m^2\phi^2$, the Fourier transforms of the propagator components are
\begin {mathletters}
\begin{eqnarray}
  \tilde G_{11}(p) & = & {1\over p_0^2-\omega_\p^2+i\eps} - {i\pi\over\omega_\p}
    \left[
	n_\p\,\delta(p_0{-}\omega_\p) + n_{-\p}\,\delta(p_0{+}\omega_\p)
    \right] \,,\\
  \tilde G_{22}(p) &=& {-1\over p_0^2-\omega_\p^2-i\eps} - {i\pi\over\omega_\p}
    \left[
    n_\p\,\delta(p_0{-}\omega_\p) + n_{-\p}\,\delta(p_0{+}\omega_\p)
    \right] \,,\\
  \tilde G_{12}(p) & = & -{i\pi\over\omega_\p}
  \left[
      n_\p\, \delta(p_0{-}\omega_{-\p}) + (1{+}n_{-\p})\,\delta(p_0{+}\omega_\p)
  \right] \,,\\
  \tilde G_{21}(p) & = & -{i\pi\over\omega_\p}
  \left[
      (1{+}n_\p) \,\delta(p_0{-}\omega_{-\p}) + n_{-\p}\,\delta(p_0{+}\omega_\p)
  \right] \,,
\end{eqnarray}%
\label{app:G-expl}%
\end {mathletters}
where $n_\p$ is the occupation number.  The retarded and advanced
propagators are
\begin{equation}
  \tilde G_R(p) = {1\over (p_0+i\epsilon)^2 -\omega_\p^2} \,,\qquad
  \tilde G_A(p) = {1\over (p_0-i\epsilon)^2 -\omega_\p^2} \,.
\end{equation}

To compute the propagator for an interacting scalar theory, it is useful to
introduce the self-energy $\Sigma(x,y)$ which is related to the
propagator by the equations
\begin{eqnarray}
  (-\partial_x^2-m^2) \,G_C(x,y) & = &
  \eta_x \, \delta(x{-}y) + \int_C\! dz\> \Sigma_C(x,z)
    \,G_C(z,y)\label{app:SD1}\,, \\
  (-\partial_y^2-m^2) \,G_C(x,y) & = &
  \eta_y \, \delta(x{-}y) + \int_C\! dz\> G_C(x,z)
    \,\Sigma_C(z,y) \,, \label{app:SD2}
\end{eqnarray}
where $\eta_x$ equals $+1$ if $x$ is on the upper part of the contour,
and $-1$ when $x$ is on the lower part.
As with the propagator, the self-energy may be decomposed into 4 components,
$\Sigma_{11}$, $\Sigma_{12}$, $\Sigma_{21}$, and $\Sigma_{22}$.
If one forms $2\times2$ matrices from the components of $G_C$ and $\Sigma_C$,
\begin {equation}
    G(x,y) \equiv
    \left(
	\begin {array}{ll} G_{11}(x,y) & G_{12}(x,y) \\
			   G_{21}(x,y) & G_{22}(x,y) \end{array}
    \right),
    \qquad
    \Sigma(x,y) \equiv
    \left(
	\begin {array}{ll} \Sigma_{11}(x,y) & \Sigma_{12}(x,y) \\
			   \Sigma_{21}(x,y) & \Sigma_{22}(x,y) \end{array}
    \right),
\end {equation}
then Eqs.\ (\ref{app:SD1}) and (\ref {app:SD2}) become
\begin{eqnarray}
    (-\partial_x^2-m^2) \, G(x,y)
    &=&
    \sigma_3 \, \delta(x{-}y)
    +
    \int dz \> \Sigma(x,z) \, \sigma_3 \, G(z,y) \,,
  \label{app:GS12}
\\
    (-\partial_y^2-m^2) \, G(x,y)
    &=&
    \sigma_3 \, \delta(x{-}y)
    +
    \int dz \> G(x,z) \, \sigma_3 \, \Sigma(z,y) \,,
  \label{app:GS21}
\end{eqnarray}
where $\int dz$ now means ordinary spacetime integration and
$\sigma_3$ is the usual Pauli matrix (but has nothing to do with spin here).
Using Eq.\ (\ref{app:GS12}) [or (\ref {app:GS21})] one can show that
the identity (\ref {app:G-id}) for the propagator implies
a corresponding identity for the self-energy,
\begin{equation}
  \Sigma_{11}+\Sigma_{22}=\Sigma_{12}+\Sigma_{21} \,.
  \label{app:Sigma-id}
\end{equation}

One can also introduce retarded and advanced self-energies
in a manner similar to Eq.~(\ref{app:GRA-def}),
\begin{eqnarray}
  \Sigma_R & = & \Sigma_{11}-\Sigma_{12} = \Sigma_{21}-\Sigma_{22}
     \nonumber \\
  \Sigma_A & = & \Sigma_{12}-\Sigma_{22} = \Sigma_{11}-\Sigma_{21}
     \label{app:SRA-def}
\end{eqnarray}
It can be easily shown that $G_{R,A}$ are related to $\Sigma_{R,A}$ by
$G_{R,A}=(G_0^{-1}-\Sigma_{R,A})^{-1}$, where $G_0$ is the corresponding
free retarded or advanced propagator.

Let us now turn to the derivation of the Boltzmann equation.  Subtracting
Eq.\ (\ref{app:SD1}) from (\ref{app:SD2}), one obtains
\begin{equation}
  (\partial_x^2-\partial_y^2) \, G_C(x,y)
  =
  \int_C\! dz\>
  \Bigl[ G_C(x,z) \, \Sigma_C(z,y) - \Sigma_C(x,z) \, G_C(z,y)\Bigr] \,.
  \label{app:dx2-dy2}
\end{equation}
Up to this point, we have not made any approximation.
Now we will assume that the
overall evolution of the system occurs on a time scale much larger than
the typical wavelength of a particle.  In terms of the propagator $G(x,y)$,
this means that it varies much more slowly as a function of the average
position ${(x+y)/2}$ than with the separation $x-y$.
Having this in mind, let us
change variables in (\ref {app:dx2-dy2})
from $x$, $y$, $z$ to new variables $X$, $s$ and $s'$, where
\begin{equation}
  x = X + {s\over2}, \qquad y = X - {s\over2}, \qquad
  z = X + {s\over2} - s' \,.
\end{equation}
Eq.\ (\ref{app:dx2-dy2}) becomes
\begin{eqnarray}
  2{\partial\over\partial X^\mu}{\partial\over\partial s_\mu} \, G_C(X,s)
  &=& \int_C\!ds'\, \left[
  G_C\biggl(X{+}{s{-}s'\over2}, s'\biggr) \,
  \Sigma_C\biggl(X{-}{s'\over2}, s{-}s'\biggr)
  \right.
\nonumber\\ && \qquad {}
    -
    \left.
  \Sigma_C\biggl(X{+}{s{-}s'\over2}, s'\biggr) \,
  G_C\biggl(X{-}{s'\over2}, s{-}s'\biggr)
  \right] .
  \label{app:dx2-dy2-1}
\end{eqnarray}
Since $G$ and $\Sigma$ vary slowly as a function of $X$, one can replace
the first argument in $G$ and $\Sigma$ on the right-hand side
of Eq.\ (\ref{app:dx2-dy2-1}) by $X$.
The (12) component of Eq.\ (\ref{app:dx2-dy2-1}) then reads,
\begin{eqnarray}
  2{\partial\over\partial X^\mu}{\partial\over\partial s_\mu} \,
  G_{12}(X,s) & = & \int\!ds'\> \Bigl[
  G_{11}(X,s')\,\Sigma_{12}(X,s{-}s') - G_{12}(X,s')\,\Sigma_{22}(X,s{-}s')
  \nonumber \\ && \quad\;{}
  -
  \Sigma_{11}(X,s')\,G_{12}(X,s{-}s') + \Sigma_{12}(X,s')\,G_{22}(X,s{-}s')
  \Bigr] \,.
  \label{app:dx2-dy2-comp}
\end{eqnarray}
Fourier transforming with respect to the relative separation,
$\tilde G(X,p)\equiv\int\!ds\>e^{-ips} \, G(X,s)$, {\em etc.},
converts Eq.~(\ref{app:dx2-dy2-comp}) to
\begin{eqnarray}
  -2ip^\mu\partial_\mu \tilde G_{12}
  &=&
  \tilde \Sigma_{12}\, (\tilde G_{11}+\tilde G_{22})
  -
  \tilde G_{12} \, (\tilde \Sigma_{11}+ \tilde \Sigma_{22}) \,,
  \label{app:before-id}
\end{eqnarray}
where, for the simplicity of notation, we have omitted the
arguments of $\tilde G$ and $\tilde \Sigma$ which are now always $(X,p)$.
Making use
of Eqs.\ (\ref{app:G-id}) and (\ref{app:Sigma-id})
allows this result to be written in the form
\begin{equation}
  -2i p^\mu\partial_\mu \, \tilde G_{12}
  =
  \tilde \Sigma_{12}\,\tilde G_{21} - \tilde G_{12} \,\tilde \Sigma_{21} \,.
  \label{app:SG-GS}
\end{equation}

\begin {figure}
\vbox
   {%
   \begin {center}
      \leavevmode
      
      \epsfbox {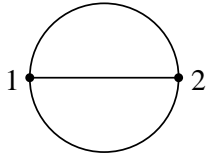}
   \end {center}
   \caption
       {%
          The leading contribution to $\Sigma_{12}$.
       \label{fig:app:scalar}
       }%
   }%
\end {figure}

Finally,
to obtain the Boltzmann equation from Eq.\ (\ref{app:SG-GS}), we make the
following ansatz for the propagators $G_{12}$ and $G_{21}$,%
\footnote
    {%
    We are assuming that the relevant particle densities are always
    low enough so that the elementary excitations of the non-equilibrium
    system are adequately viewed as single fundamental particles.
    See \cite {Jeon&Yaffe} for a discussion of the construction
    of an ``effective'' kinetic theory which remains valid
    when this assumption does not hold.
    }
\begin{eqnarray}
  \tilde G_{12}(X,p) & = &
  -{i\pi\over\omega_\p} \biggl[\,\delta(p_0{-}\omega_\p) \, n_\p(X) +
    \delta(p_0{+}\omega_\p) \, (1{+}n_{-\p}(X)) \biggr] \,,\\
  \tilde G_{21}(X,p) & = &
  -{i\pi\over\omega_\p} \biggl[\,\delta(p_0{-}\omega_\p) \, (1{+}n_\p(X)) +
    \delta(p_0{+}\omega_\p) \, n_{-\p}(X) \biggr] \,.
\label {app:G-ansatz}
\end{eqnarray}
In other words, one assumes that $\tilde G$ has the same form as the free
propagator, except that the distribution function is now an
arbitrary function of both $X$ and $\p$.  By comparing
the coefficient of $\delta(p_0-\omega_\p)$, one derives from Eq.\
(\ref{app:SG-GS}) that
\begin{equation}
  (\partial_t + \v\cdot \partial_\x) \, n_\p
  = {i\over2\omega_\p} \, \Bigl[\,
  \tilde\Sigma_{12}(\omega_\p,\p) \, (1{+}n_\p) -
  \tilde\Sigma_{21}(\omega_\p,\p) \, n_\p
  \Bigr] \,.
  \label{app:Boltzmann-prel}
\end{equation}
One can see the Boltzmann equation emerging. Indeed, the term
${i\over2\omega_\p}\tilde\Sigma_{12}(1{+}n_\p)$ is the ``gain'' term and
$-{i\over2\omega_\p}\tilde\Sigma_{21}\,n_\p$ is the ``loss'' term in the
collision integral.  To produce the conventional form of the
Boltzmann equation, we need to compute the leading-order contribution to
self-energy.
There is no one-loop contribution to $\tilde\Sigma_{12}$ or $\tilde\Sigma_{21}$.
The first non-zero contribution comes from the two-loop
diagram (Fig.\ \ref{fig:app:scalar}).  Using the explicit propagators
(\ref{app:G-expl}), one finds,
\begin{eqnarray}
  \tilde\Sigma_{12}(\omega_\p,\p) & = &
  -{i\lambda^2\over2}\int\!{d\p'd\k \over
  (2\pi)^6 \, 2\omega_{\p'} 2\omega_{\k} 2\omega_{\k'}} \>
  n_{\p'} \, n_{\k} \, (1{+}n_{\k'}) \,
  2\pi\delta(\omega_{\p}{+}\omega_{\p'}{-}\omega_{\k}{-}\omega_{\k'}) \,,
  \nonumber\\
  \tilde\Sigma_{21}(\omega_\p,\p) & = &
  -{i\lambda^2\over2}\int\!{d\p'd\k \over
  (2\pi)^6 \, 2\omega_{\p'} 2\omega_{\k} 2\omega_{\k'}} \>
  (1{+}n_{\p'}) \, (1{+}n_{\k}) \, n_{\k'} \,
  2\pi\delta(\omega_{\p}{+}\omega_{\p'}{-}\omega_{\k}{-}\omega_{\k'}) \,,
  \label{app:S12-S21}
\end{eqnarray}
where $\k'\equiv\p+\p'-\k$.  Substituting Eq.\ (\ref{app:S12-S21}) into Eq.\
(\ref{app:Boltzmann-prel}), one obtains the Boltzmann equation
\begin{eqnarray}
  (\partial_t + \v \cdot \partial_\x) \, n_\p & = &{\lambda^2\over2}
  \int\!{d\p'\,d\k\over (2\pi)^6 \, 2\omega_\p 2\omega_{\p'}
  2\omega_\k 2\omega_{k'}} \>
  2\pi\delta(\omega_\p{+}\omega_{\p'}{-}\omega_\k{-}\omega_{\k'})
  \nonumber\\ && \quad {} \times
  \Bigl[n_{\p_1} \, n_{\p_2} \, (1{+}n_{\p_3}) \, (1{+}n_\p)
  - (1{+}n_{\p_1}) \, (1{+}n_{\p_2}) \, n_{\p_3} \, n_\p \Bigr] \,,
  \label{app:Boltzmann}
\end{eqnarray}
which coincides with the result one would derive
naively from kinetic theory.

\subsection{The multi-component case}

The extension of the derivation in the previous subsection to the case of
a multi-component scalar field is straightforward.  Let the field be
$\phi^a$, where $a$ is some isospin index.
The propagator and self-energy components become matrices,
$G_{12} = ||G_{12}^{ab}||$, {\em etc.},
but everything in the preceding discussion up to (and including)
Eq.~(\ref {app:dx2-dy2-comp}) remains valid.
However,
Eq.\ (\ref{app:before-id}) is not correct
since the components of $G$ and $\Sigma$ no longer commute.
Instead, one rewrites Eq.\ (\ref{app:dx2-dy2-comp}) as
\begin{equation}
  (\partial_x^2-\partial_y^2) \, G_{12}
  = {1\over2} \biggl( \{ G_{11}{+}G_{22}, \Sigma_{12} \} +
    [ G_{11}{-}G_{22}, \Sigma_{12} ] -
    \{ \Sigma_{11}{+}\Sigma_{22}, G_{12} \}
   -[ \Sigma_{11}{-}\Sigma_{22}, G_{12} ]\biggr).
\end{equation}
The anti-commutator terms can be simplified by using Eqs.\
(\ref{app:G-id})  and (\ref{app:Sigma-id})
to produce
\begin{equation}
  (\partial_x^2-\partial_y^2) \, G_{12}=
  {1\over2} \biggl( \{ G_{21}, \Sigma_{12} \} - \{ \Sigma_{21}, G_{12} \}
  + [ G_{11}{-}G_{22}, \Sigma_{12} ] - [ \Sigma_{11}{-}\Sigma_{22}, G_{12} ]
  \biggr) \,.
  \label{app:comm-anticomm}
\end{equation}
To progress further, we make the following ansatz for the propagator
\begin{eqnarray}
  G^{ab}_{12}(X,p) & = &
  -{i\pi\over\omega_\p} \left[\delta(p_0{-}\omega_\p) \, n^{ab}_\p(X)
      + \delta(p_0{+}\omega_\p) \, (\delta^{ab}+n^{ba}_{-\p}(X))\right] ,
  \nonumber\\
  G^{ab}_{21}(X,p) & = &
  -{i\pi\over\omega_\p}
  \left[\delta(p_0{-}\omega_\p) \,  (\delta^{ab}+n^{ab}_\p(X))
      + \delta(p_0{+}\omega_\p) \, n^{ba}_{-\p}(X)\right] .
  \label{app:G-ansatz-ab}
\end{eqnarray}
Physically, $n^{ab}_\p(X)$ is the density matrix (in isospin space) of the
momentum $\p$ excitations.
The ansatz (\ref{app:G-ansatz-ab}) immediately implies that
$G_{12}{-}G_{21}$ is proportional to $\delta^{ab}$,
and this in turn implies the same for $G_{11}{-}G_{22}$.
Therefore, the first commutator term in (\ref {app:comm-anticomm}) vanishes
and the kinetic equation becomes
\begin{eqnarray}
  (\partial_t+\v\cdot\partial_\x) \, n_\p &=& {i\over4\omega_\p}\biggl(
  \{ \Sigma_{12}(\omega_\p,\p), (1+n_\p) \}
  - \{ \Sigma_{21}(\omega_\p,\p), n_\p \}
\nonumber\\&&\kern 0.5in{}
  - [\Sigma_{11}(\omega_\p, \p){-}\Sigma_{22}(\omega_\p, \p), n_\p]\biggr) \,.
  \label{app:Sn-nS-ab}
\end{eqnarray}
This has the same form Eq.\ (\ref{eq:C}) in the main
text if one identifies
\begin{equation}
  \I_- = {i\over2\omega_\p} \, \Sigma_{12}(\omega_\p,\p) \,, \qquad
  \I_+ = {i\over2\omega_\p} \, \Sigma_{21}(\omega_\p,\p) \,,
\end{equation}
and
\begin{equation}
  \text{Re}\overline{\Sigma} ={1\over4\omega_\p} \,(\Sigma_{11}{-}\Sigma_{22})
  = {1\over4\omega_\p} \, (\Sigma_R{+}\Sigma_A)
  = {1\over2\omega_\p} \, \text{Re}\,\Sigma_R(\omega_\p, \p) \,.
\end{equation}
The quantity
$\overline{\Sigma}\equiv{1\over2\omega_\p}\Sigma_R$ has a simple physical
meaning: it is the correction to the energy of an excitation with momentum $\p$
(in other words, it is the self-energy in the non-relativistic normalization.)
It is straightforward to find the explicit form of $\I_{\pm}$
at the leading (two-loop) level, and show that one obtains the
equations (\ref{eq:I-}) and (\ref{eq:I+}) discussed in the main text.
Let us, however, move on to the case of gauge theories.

\subsection{Gauge theories}

The approach of the previous subsection can be carried over to the
gauge theory case without substantial modification.
We consider scalar QED first, and will find the Boltzmann equation
describing the kinetics of the hard scalar particles.  The basic equation
remains Eq.\ (\ref{app:Boltzmann-prel}).  However, the computation of the
scalar self-energy $\Sigma$ is slightly more complicated than in the
$\phi^4$ case.  For scalar QED, the leading contribution to the self-energy
$\Sigma_{12}$ comes from the one-loop diagram:
$$
      
      \epsfbox {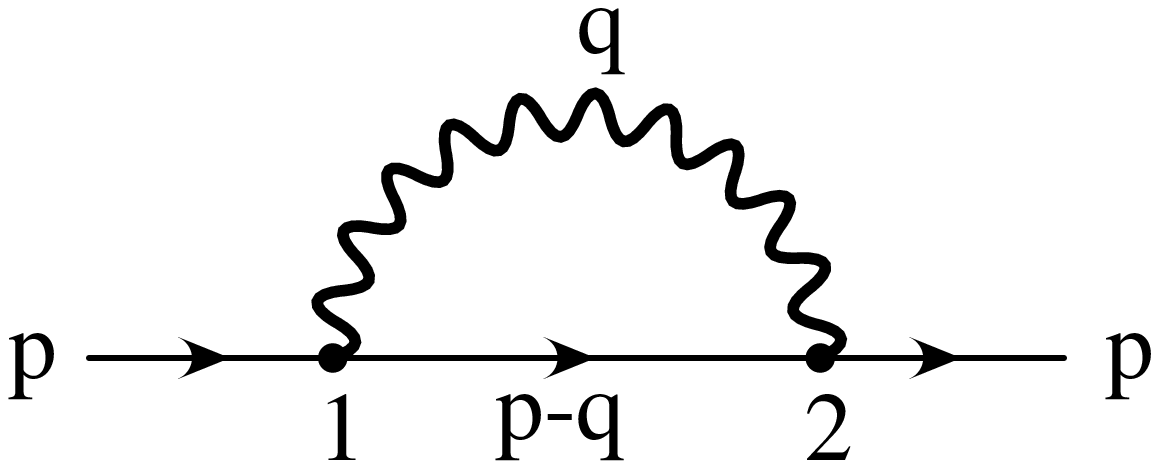}
$$
This gives
\begin{equation}
  \tilde\Sigma_{12}(p) = -ie^2\int\!{d^4q\over(2\pi)^4} \>
  (2p{-}q)_\mu (2p{-}q)_\nu \, D^{\mu\nu}_{12}(q) \, \tilde G_{12}(p{-}q)
\label {app:S12}
\end{equation}
where $D^{\mu\nu}$ is the photon propagator.  Typical scatterings between bosons
have small momentum exchange, so we will assume that the
internal photon momentum $q$ is small.  The soft photon propagator $D(q)$ is
obtained by summing the bubble diagrams:
$$
      
      \epsfbox{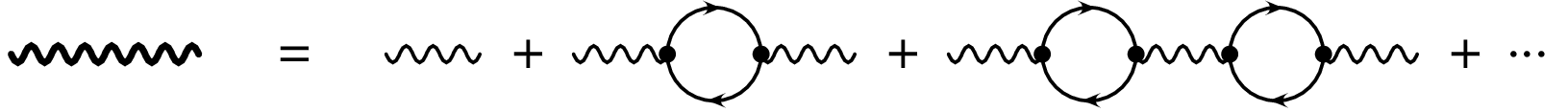}
$$
or equivalently, introducing the photon self-energy $\Pi$.
$\Pi$ has
four components and is related to the photon propagator through the equations
[see (\ref{app:GS12}) with $\Sigma \to \Pi$ and $m=0$]
\begin{eqnarray}
  q^2 D_{12} & = & \Pi_{11}\,D_{12} - \Pi_{12}\,D_{22} \,, \nonumber\\
  q^2 D_{22} & = & \Pi_{21}\,D_{12} - \Pi_{22}\,D_{22} - 1 \,.\label{app:PiD}
\end{eqnarray}
We have suppressed the Lorentz indices here
for notational simplicity.
in Eq.\ (\ref{app:PiD}).
[In fact, the longitudinal and transverse parts of $D$ and
$\Pi$ satisfy Eqs.\ (\ref{app:PiD}) separately.]  Solving for $D_{12}$,
one finds
\begin{equation}
  D_{12}(q) =  { \Pi_{12} \over
  (q^2-\Pi_{11})(q^2+\Pi_{22})+\Pi_{12}\,\Pi_{21} } \,.
\end{equation}
As in the scalar case, the photon self-energy $\Pi$ satisfies the
identity (\ref{app:Sigma-id}) and the retarded and advanced self-energies
$\Pi^{R,A}$ can be introduced in a manner similar to Eq.\
(\ref{app:SRA-def}). It is easy to show that
$(q^2-\Pi_{11})(q^2+\Pi_{22})+\Pi_{12}\,\Pi_{21} = (q^2-\Pi_R)(q^2-\Pi_A)$,
and therefore
\begin{equation}
  D_{12}(q) = {\Pi_{12} \over (q^2-\Pi_R)(q^2-\Pi_A) }
            =  D_R \, \Pi_{12} \, D_A =  |D_R|^2 \, \Pi_{12} \,.
\label {app:D12}
\end{equation}
The photon self-energy $\Pi_{12}$ is determined by the scalar one-loop diagram,
which gives
\begin{equation}
  \Pi^{\mu\nu}_{12}(q) = -ie^2\int\!{dp'\over(2\pi)^4} \>
  (2p'{+}q)^\mu (2p'{+}q)^\nu \, G_{12}(p') \, G_{21}(p'{+}q) \,.
  \label{app:qed-pi}
\end{equation}
We now insert the ansatz (analogous to (\ref {app:G-ansatz}))
for the propagator of the complex scalar $\phi$,
\begin{equation}
  \tilde G_{12}(p) = -{i\pi\over\omega_\p} \,
  \Bigl[\,
      n_\p \, \delta(p_0{-}\omega_\p)
      + (1{+}\bar{n}_{-\p}) \,\delta(p_0{+}\omega_\p)
  \Bigr] \,.
  \label{app:qed-ansatz}
\end{equation}
Here, $n_\p$ and $\bar{n}_\p$ are distribution functions of particles and
anti-particles, respectively.  Substituting this into
Eq.\ (\ref{app:qed-pi}), one finds
\begin{equation}
  \Pi^{\mu\nu}_{12}(q) = ie^2\int\!{d\p'\over(2\pi)^3} \>
  (2p'+q)^\mu (2p'+q)^\nu \,
  \left[n_{\p'+\q} \, (1{+}n_{\p'}) + \bar{n}_{\p'{+}\q}\, (1{+}\bar{n}_{\p'})
  \right] \,.
\label {app:pimunu}
\end{equation}
Inserting (\ref {app:D12}), (\ref {app:qed-ansatz}), and (\ref {app:pimunu})
in the scalar self-energy (\ref {app:S12}) yields
\begin{eqnarray}
  \Sigma_{12}(\omega_\p,\p) &=&
  -ie^4 \int\! {d\p'\, d\q\over (2\pi)^6 \, 2\omega_{\p'}
  2\omega_{\p-\q} 2\omega_{\p'+\q}} \>
  (2\pi)\delta(\omega_\p{+}\omega_{\p'}{-}\omega_{\p-\q}{-}\omega_{\p'+\q})
  \nonumber\\
  &\times&
  \left|(2p{-}q)_\mu (2p'{+}q)_\nu D^{\mu\nu}_R(q)\right|^2 \,
  \left[(1{+}n_{\p'}) \, n_{\p-\q} \, n_{\p'+\q} +
  (1{+}\bar{n}_{\p'}) \, n_{\p-\q} \, \bar{n}_{\p'+\q}\right].
  \label{app:qed:Sigman3}
\end{eqnarray}
The other off-diagonal self-energy component,
$\Sigma_{21}$, may be computed completely analogously.
Inserting the expressions for $\Sigma_{12}$ and $\Sigma_{21}$
into Eq.~(\ref {app:Boltzmann-prel})
yields the scalar QED Boltzmann equation,
\begin{eqnarray}
  (\partial_t+\v\cdot\partial_\x) \, n_{\p} & = &
  \int\!{d\p'\, d\k \over (2\pi)^6 \,
  2\omega_p 2\omega_{\p'} 2\omega_\k 2\omega_\k} \>
  |M_{\p\p'\to\k\k'}|^2 \,
  (2\pi)\delta(\omega_\p{+}\omega_{\p'}{-}\omega_\k{-}\omega_{\k'})
  \nonumber\\ && \qquad {} \times
  \biggl\{ (1{+}n_\p) \, n_\k \, \left[(1{+}n_{\p'}) \, n_{\k'} +
  (1{+}\bar{n}_{\p'}) \, \bar{n}_{\k'} \right]
  \nonumber\\ && \quad\qquad {}
  - n_\p \, (1{+}n_\k) \, \left[ n_{\p'} \, (1{+}n_{\k'}) +
  \bar{n}_{\p'} \, (1{+}\bar{n}_{\k'}) \right] \biggr\}.
\end{eqnarray}
where we have introduced $k\equiv p-q$, $k'\equiv p'+q$,
and the scattering amplitude $M_{\p\p'\to\k\k'} \equiv e^2(p{+}p')_\mu
(k{+}k')_\nu D^{\mu\nu}_R(q)$.
This has the same form as the naive Boltzmann equation, with the
exception that the scattering amplitude is to be computed
using the resummed propagator $D_R(q)$ for the exchanged photon
instead of the bare photon propagator.

Finally,
combining our treatment of multi-component scalar theory
with that of QED, one may write down the Boltzmann equation for a
non-Abelian gauge theory.  The distribution functions become matrices
$n^{a\bar b}_\k$ with respect to group indices.%
\footnote{To simplify the discussion,
we assume that distribution functions are trivial with
respect to polarization.}
The Boltzmann equation has the form shown in Eq.~(\ref{app:Sn-nS-ab}).
The loss and gain terms $\I_-$ and $\I_+$, which come from
one-loop contributions to the self-energy
(computed with a soft resummed gauge boson propagator),
are trivial generalizations of Eq.~(\ref{app:qed:Sigman3}).
The final results are given in Eqs.~(\ref{eq:boltz})--(\ref{eq:C})
of the main text.



\begin {references}

\bibitem {reviews}
   A. Cohen, D. Kaplan and A. Nelson,
   {\sl Annu.\ Rev.\ Nucl.\ Part.\ Sci.}\ {\bf 43}, 27 (1993);
   V. Rubakov and M. Shaposhnikov,
   {\tt hep-ph/9603208},
   {\sl Usp.\ Fiz.\ Nauk} {\bf 166}, 493 (1996)
   [{\sl Phys.\ Usp.}\ {\bf 39}, 461 (1996)];
   M. Trodden,
   {\tt hep-ph/9803479},
   Case Western report no.\ CWRU-P6-98.

\bibitem {notransition}
   K. Kajantie, M. Laine, K. Rummukainen, and M. Shaposhnikov,
   {\tt hep-ph/9605288},
   {\sl Phys.\ Rev.\ Lett.}\ {\bf 77}, 2887 (1996).
   S. Elitzur,
   Phys.\ Rev.\ {\bf D12}, 3978 (1975).


\bibitem {ASY1}
  P. Arnold, D. Son, and L. Yaffe,
  {\tt hep-ph/9609481},
  {\sl Phys.\ Rev.}\ {\bf D55}, 6264 (1997).

\bibitem {Huet&Son}
    P. Huet and D. Son,
    {\tt hep-ph/9610259},
    {\sl Phys.\ Lett.}\ {\bf B393}, 94 (1997);
    D. Son,
    {\tt hep-ph/9707351},
    Univ.\ Washington report no.\ UW-PT-97-19.

\bibitem{Arnold}
    P. Arnold,
    {\tt hep-ph/9701393},
    {\sl Phys.\ Rev.}\ {\bf D55}, 7781 (1997).

\bibitem{Bodeker}
    D. B\"odeker,
    {\tt hep-ph/9810430},
    {\sl Phys.\ Lett.}\ {\bf B426}, 351 (1998).

\bibitem{scale}
    T. Appelquist and R. Pisarski,
      {\sl Phys.\ Rev.}\ {\bf D23}, 2305 (1981);
    S. Nadkarni,
      {\sl Phys.\ Rev.}\ {\bf D27}, 917 (1983);
      {\sl Phys.\ Rev.}\ {\bf D38}, 3287 (1988);
      {\sl Phys.\ Rev.\ Lett.}\ {\bf 60}, 491 (1988);
    N. Landsman,
      {\sl Nucl.\ Phys.}\ {\bf B322}, 498 (1989);
    K. Farakos, K. Kajantie, M. Shaposhnikov,
      {\sl Nucl.\ Phys.}\ {\bf B425}, 67 (1994).

\bibitem{Moore}
    Guy Moore, private communication.

\bibitem{mess}
  D. Bodeker, L. McLerran, and A. Smilga,
  {\tt hep-th/9504123},
  {\sl Phys.~Rev.} {\bf D52}, 4675 (1995).

\bibitem{classical}
   D. Grigoriev and V. Rubakov,
   Nucl.\ Phys.\ {\bf B299}, 671 (1988);
   D. Grigoriev, V. Rubakov, and M. Shaposhnikov,
   Nucl.\ Phys.\ {\bf B308}, 885 (1988);
   for a review, see
   E. Iancu,
   {\tt hep-ph/9807299},
   Saclay report no.\ SACLAY-T98-073.

\bibitem{ZinnJustin}
    J. Zinn-Justin, {\sl Quantum Field Theory and Critical Phenomena},
    2nd edition (Oxford University Press, 1993).

\bibitem{paper2}
    P. Arnold, D. Son, and L. Yaffe,
    Univ.\ of Washington report no.\ UW/PT 98-17, in preparation.

\bibitem{Selikhov}
    A. Selikhov and M. Gyulassy,
    {\tt nucl-ph/9307007},
    {\sl Phys.\ Lett.}\ {\bf B316}, 373 (1993).

\bibitem{gammag}
    R. Pisarski,
    {\sl Phys.\ Rev.\ Lett.}\ {\bf 63}, 1129 (1989);
    {\sl Phys. Rev.}\ {\bf D47}, 5589 (1993).

\bibitem{jj}
   P. Arnold and L. Yaffe,
   {\tt hep-ph/9709449},
   {\sl Phys.\ Rev.}\ {\bf D57}, 1178 (1998).

\bibitem{taularge}
   A. Hosoya and K. Kajantie,
   {\sl Nucl.\ Phys.}\ {\bf B250}, 666 (1985);
   G. Baym, H. Monien, C. Pethick, D. Ravenhall,
   {\sl Phys.\ Rev.\ Lett.}\ {\bf 64}, 1867 (1990).

\bibitem{smilga}
  V. Lebedev and A. Smilga,
  {\sl Physica} {\bf A181}, 187 (1992).

\bibitem{BI2}
   J. Blaizot and E. Iancu,
   {\sl Phys.\ Rev.\ Lett.}\ {\bf 76}, 3080 (1996);
   {\sl Phys.\ Rev.}\ {\bf D55}, 973 (1997); {\bf D56}, 7877.

\bibitem{Jeon}
   S. Jeon,
   {\sl Phys. Rev.}\ {\bf D52}, 3591 (1995).

\bibitem {Jeon&Yaffe}
   S. Jeon and L. Yaffe,
   {\sl Phys. Rev.}\ {\bf D53}, 5799 (1996).

\bibitem{Heiselberg}
   H. Heiselberg,
   {\tt hep-ph/9401317},
   {\sl Phys.\ Rev.\ Lett.}\ {\bf 72}, 3013 (1994).

\bibitem{Selikhov2a}
   A. Selikhov and M. Gyulassy,
   {\sl Phys.\ Rev.}\ {\bf C49}, 1726 (1998).

\bibitem{Selikhov2b}
   A. Selikhov,
   {\sl Phys.\ Lett. B} {\bf 268}, 263 (1991); {\bf 285}, 398(E) (1992).

\bibitem{Blaizot&Iancu}
   J. Blaizot and E. Iancu,
   {\sl Nucl.\ Phys.}\ {\bf B390}, 589 (1993).

\bibitem{deGroot}
   S. de Groot, W. van Leeuwen, Ch.\ van Weert,
   {\sl Relativistic Kinetic Theory} (North-Holland, 1980).

\bibitem{Snider}
   R. Snider in
   {\sl Lecture Notes in Physics}, vol. 31: {\sl Transport Phenomena}
   (Springer-verlag, 1974), ed. J. Ehlers {\it et al.}

\bibitem{adjXadj}
   S. Mr\'{o}wczy\'{n}ski,
   {\sl Phys.\ Rev.}\ {\bf D39}, 1940 (1989);
   H.-Th. Elze and U. Heinz,
   {\sl Phys.\ Rept.}\ {\bf 183}, 81 (1989);
   J. Blaizot and E. Iancu,
   {\sl Nucl.\ Phys.}\ {\bf B417}, 609 (1994);
   and references therein.

\bibitem{Botermans}
   W. Botermans and R. Malfliet,
   {\sl Phys.\ Lett.}\ {\bf B215}, 617 (1988).

\bibitem{Pi}
   H. Weldon,
     Phys.\ Rev.\ D {\bf 26}, 1394 (1982);
   U. Heinz,
     Ann.\ Phys.\ (N.Y.) {\bf 161}, 48 (1985); {\bf 168}, 148 (1986).

\bibitem{W}
   J. Blaizot and I. Iancu,
   {\sl Phys.\ Rev.\ Lett.}\ {\bf 72}, 3317 (1994).

\bibitem{HeiselbergViscosity}
   H. Heiselberg,
   {\sl Phys.\ Rev.}\ {\bf D49}, 4739 (1994).

\bibitem{ZZ}
   J. Zinn-Justin and D. Zwanziger,
   {\sl Nucl.\ Phys.}\ {\bf B295} [FS21], 297 (1987).

\bibitem{Schwinger}
   J. Schwinger, J. Math. Phys. {\bf 2}, 407 (1961).

\bibitem{Keldysh}
   L.V. Keldysh, Zh. Eksp. Teor. Fiz. {\bf 47}, 1515 (1964)
   (Sov.  Phys. JETP {\bf 20}, 1018 (1965)).

\bibitem{LifshitzPitaevskii}
   E.M. Lifshitz, L.P. Pitaevskii, {\sl Physical Kinetics} (Pergamon
   Press, 1981).


\end {references}
\end {document}